\documentclass{sig-alternate}
\usepackage{mathptmx} 

\usepackage{fancyhdr}
\usepackage[keeplastbox]{flushend}

\usepackage{amsmath,amssymb,amsfonts}

\usepackage{algpseudocode}
\usepackage{graphicx}
\usepackage[export]{adjustbox}
\usepackage{textcomp}
\usepackage{xcolor}
\usepackage{comment}
\usepackage{authblk}
\usepackage{algorithm}

\usepackage{subcaption}

\usepackage{graphicx}

\newcommand{\etal}{\emph{et al.\ }}

\newcommand{\fixit}[1] {
  {\color{red}{#1}}}

\newcommand{\newrev}[1] {{\color{red}{#1}}}

\fancypagestyle{firstpage}{
  \fancyhf{}
  
  \fancyhead[C]{\vspace{10pt}\normalsize{
      \textbf{} }\\\vspace{-25pt}} 
  \fancyfoot[C]{\thepage}
}

\title{A Structured Method for Compiling and Optimizing  QAOA Circuits in Quantum Computing} 
\author{Yuwei Jin$^1$, Jason Luo$^1$, Lucent Fong$^1$, Yanhao Chen$^1$, \\ Ari B. Hayes$^1$, Chi Zhang$^2$, Fei Hua$^1$, Eddy Z. Zhang }

\affil[1]{Rutgers University} \affil[2]{University of Pittsburgh}

\begin{document}
\maketitle
\thispagestyle{firstpage}
\pagestyle{plain}


\begin{abstract}
\vspace{-5pt}
Quantum Approximation Optimization Algorithm (QAOA) is a highly advocated variational algorithm for solving the combinatorial optimization problem. One critical feature in the quantum circuit of QAOA algorithm is that it consists of two-qubit operators that commute. The flexibility in reordering the two-qubit gates allows compiler optimizations to generate circuits with better depths, gate count, and fidelity. However, it also imposes significant challenges due to additional freedom exposed in the compilation. Prior studies lack the following: (1) Performance guarantee, (2) Scalability, and (3) Awareness of regularity in scalable hardware.

 We propose a structured method that ensures linear depth for any compiled QAOA circuit on multi-dimensional quantum architectures. We also demonstrate how our method runs on Google Sycamore and IBM Non-linear architectures in a scalable manner and in linear time. Overall, we can compile a circuit with up to 1024 qubits in 10 seconds with a $3.8X$ speedup in depth, $17\%$ reduction in gate count, and $18X$ improvement for circuit ESP.
 

\end{abstract}

\section{Introduction}
\label{sec:intro}

Quantum computing has garnered considerable attention thanks to recent advances in quantum mechanics. The coherence time of a single qubit has increased exponentially in the past two decades \cite{oliver+:mrs13}, analogue to Moore's law for semi-conductor computers. Industry is now able to build quantum computers up to between 53 and 127 qubits. Moreover, IBM is projected to release a quantum computer with over 1,000 qubits in the near future \cite{ibm1000}.

Variational quantum algorithm (VQA) \cite{VQE2014} is an important class of quantum applications that holds promise for achieving near-term quantum supremacy. 
Today's quantum devices are prone to errors. Quantum error correction (QEC) is necessary for applications that have relatively large depths such as Shor's algorithm  \cite{shor:siamjcomput97}. But QEC is prohibitive due to the scale of today's quantum computers. For instance, to run a Shor's algorithm with reasonable size (factoring a 1,000 bit number), it will need millions of physical qubits, which is not possible at this stage.  VQA algorithms typically have modest depths and may not need QEC at this stage. Therefore it can exploit near-term quantum hardware.   

  Quantum approximate optimization algorithm (QAOA) \cite{QAOA:farhi2014quantum,QAOA:farhi2017quantum,QAOA:farhi2021quantum} is a highly advocated variational quantum algorithm. It is a general-purpose algorithm that has broad applications in combinatorial optimization, for instance, quadratic 0-1 programming and max-cut problem. We focus on the optimization of QAOA circuit in this paper.
  
  To fully explore the power of QAOA algorithms,  the QAOA circuits must be compiled to be resilient to noise. To compile and optimize a QAOA circuit, there are two approaches. The first one is the generic compiler \cite{zhang+:asplos21, Li+:ASPLOS19,murali+:asplos19,siraichi+:oopsla19,Siraichi+:CGO18,Zulehner+:DATE18,Zulehner+:ICRC17, tan+:iccad20} including  Qiskit \cite{IBMQiskit} and T$\mid$ket$>$ \cite{sivarajah+:iop20}. The second is the application-specific compiler. The application-specific compiler takes into account the flexiblity in permuting operators  in QAOA circuits, while the generic compilers do not. An example is shown in Fig.\ \ref{fig:qaoaexample}, in which after permuting the gates, the logical circuit depth reduces by 25\%. 
  

\begin{figure}[htb]
    \centering
    \includegraphics[width=0.4
    \textwidth]{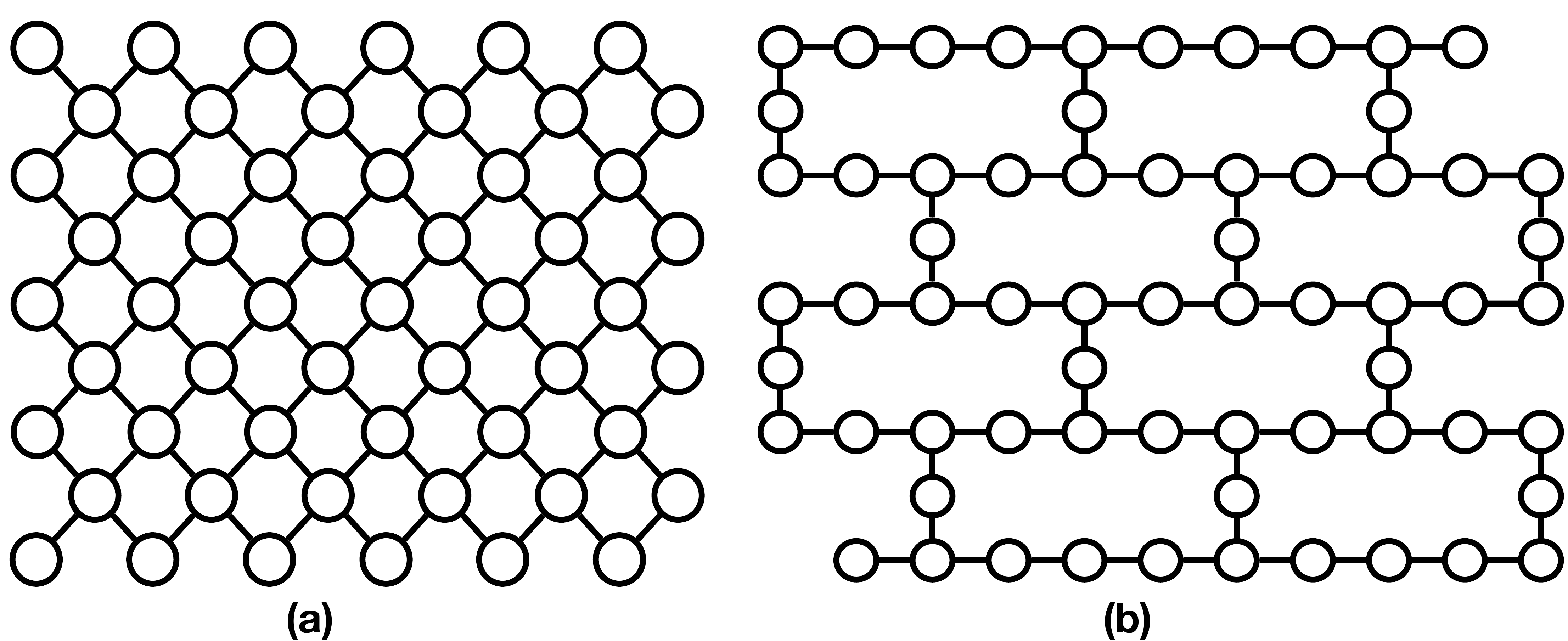}
    \caption{(a) Google Sycamore (b) IBM Heavy-hex}
    \label{fig:google_ibm}
\end{figure}

\begin{figure*}[htb]
    \centering
    \includegraphics[width=0.85
    \textwidth]{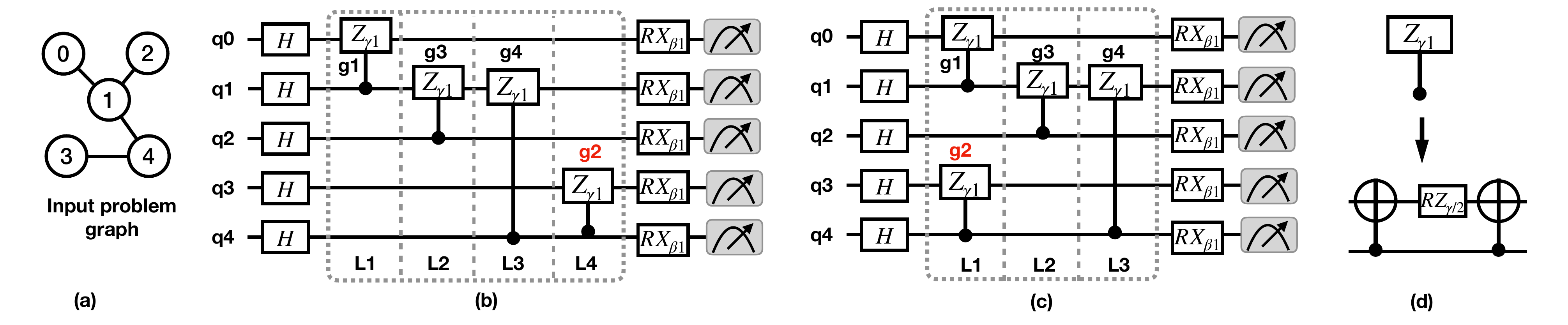}
    \caption{A five-qubit QAOA-maxcut circuit example: (a) the input problem graph where MAX-CUT is applied on. (b) and (c) are two valid circuits corresponding to  the input problem graph. (d) One CPHASE gate decomposition. 
 }
    \label{fig:qaoaexample}
\end{figure*}

The application-specific compilers have demonstrated better results than the generic compilers in terms of both performance and compilation time \cite{lao+:arxiv20, alam+:micro20}. Tan \etal \cite{tan+:iccad20} provide a constrained SAT solver to achieve the optimal compilation result for QAOA. Alam \etal \cite{alam+:micro20} use the heuristic of connectivity strength to improve the depth and fidelity of the compiled circuit. Paulihedral \cite{li+:asplos22} is a compiler framework that defines intermediate representation and optimizes the compilation of Pauli-strings in Hamiltionian simulation. 2QAN \cite{lao+:arxiv21twoqan} proposes a permutation-aware heuristic that has quadratic-time complexity and uses unitary unifying to enhance the performance.

However, existing application-specific approaches lack rigorous analytical investigation on the QAOA compilation problem.  As far as we are concerned, none of the existing compilers provides performance guarantee of the compiled circuit. The compilation of a logical circuit to a hardware circuit involves adding additional gates. If not properly compiled or optimized, a circuit might have unnecessarily large depth, sabotaging the depth benefit of QAOA algorithms. The problem exacerbates when the problem size scale to beyond 1,000 qubits. Previous studies are lacking from the following perspectives. 

\textbf{(1) Lack of scalability:} { Optimal solvers such as that by Zhang \etal \cite{zhang+:asplos21} can be adapted to compile QAOA circuits. }   However, it may not be able to handle large scale circuits. Optimal compilation is not a viable solution in the long run.

\textbf{(2) No Provable Performance Guarantee} Heuristic compilers including QAIM \cite{alam+:dac20, alam+:micro20}, Paulihedral \cite{li+:asplos22}, and 2QAN \cite{lao+:arxiv21twoqan} can compile reasonably fast for circuits with hundreds of qubits. However, it is not clear how far the compiled circuit is from the optimal result, given any of the existing approaches. Based on our experiment results, when going beyond 256 qubits, QAIM runs for more than {two hours}.  Paulihedral can run fast but provide less than desired compilation results for large-scale graphs. Using greedy heuristics can help solve the compilation problem in the near term, but we need to have an in-depth understanding of a given core heuristic as well as its application scenario.

\textbf{(3) Obliviousness to Regularity in the Scalable Quantum Hardware} Neither optimal compilers \cite{zhang+:asplos21} or heuristic compilers \cite{li+:asplos22, lao+:arxiv21twoqan, alam+:dac20, alam+:micro20} take into account of the regular structure in today's quantum hardware that scale. We show Google Sycamore structure in Fig.\ref{fig:google_ibm}(a)  where an obvious pattern of diagonal links between qubits in a 2D lattice is manifested. Similarly recent IBM architectures show regularity in their structure: they use the { heavy-hex architecture \cite{ibm_heavy_hex}}. IBM architecture is shown in Fig.\ \ref{fig:google_ibm}(b). 

To this end, we propose a structured method for the compilation and optimization of QAOA circuits. Our method provides scalability,  performance guarantee, and regularity-exploitation in scalable quantum hardware. Our contributions are summarized as follows:
\vspace{-5pt}
\begin{itemize}
    \item {\textbf{Linear depth {for {multi-dimensional}
    {architecture:}}} Our compiled circuit is guaranteed to provide linear depth for Google Sycamore, IBM heavy-hex, and grid.} Specifically, it guarantees   $\frac{5n}{2}$ depth  for Google Sycamore architecture, \newrev{$6n$} for IBM heavy-hex architecture, and $\frac{3n}{2}$ depth for grid (if we ignore the sub-linear terms), in worst case scenario, where $n$ is the number of qubits. 
    \vspace{-5pt}
    \item \textbf{Hardware regularity exploitation:} { We exploit the hardware regularity through the \emph{multi-dimensional all-to-all permutation and swaps (Maaps)} method we propose in this paper. The Maaps method provides a new way of thinking on achieving the all-to-all interaction in multi-dimensional architectures. We used Maaps to show how to come up with solutions for Google Sycamore, grid, and an additional imaginary hexagon architecture. }    
    \vspace{-5pt}
    \item \textbf{Enabling scalability:} Our compilation does not require an optimal solver to achieve a desired depth. It has linear time complexity. Hence it can compile large circuits very fast. It can run within 9  seconds for a circuit with 1,024 qubits. The compile time scales linearly with the number of qubits in quantum hardware, {and yet the compiler provides linear depth guarantee.}
    \vspace{-5pt}
    \item \textbf{Application beyond QAOA:} Our compiler framework can be applied to 2-local Hamiltonian simulation in addition to QAOA. The compilation and optimization of QAOA is essentially the same as that of 2-local Hamiltonian simulation. In both cases, a \emph{problem-graph} determines the structure of the logical circuit, and the 2-qubit operators can permute.  2-local Hamiltonian is a common type of Hamiltonian simulation that originate in many physical systems, such as the traversing Ising model, XY model, and the Heiserburg model. 
    
\end{itemize}

Overall we provide a simple yet effective compiler framework for compiling QAOA and 2-local Hamitonian simulation circuits. Compared with QAIM \cite{alam+:dac20}, our experiments show that we can achieve up to 3.8X speedup for depth, 17\% reduction in gate count, and 18X in fidelity improvement. In the meantime, our compiler has low overhead. We have evaluated circuits with up to 1,024 qubits in this paper and the compiler takes no more than 5 minutes, often running in less than 10 seconds, while prior study may take hours to compile a circuit with $>256$ qubits.

{We recognize that an independent study by Weindenfeller \etal \cite{weidenfeller+:arxiv22} comes up with linear solutions for grid and IBM heavy-hex architectures at nearly the same time as we do (February 2022 v.s. ours first 2xN grid solution in December 2021 and NxN grid solution in April 2022). Their solution for grid is similar to ours, except that our solution reduces 25\% depth compared with theirs when SWAP is not implemented as three CNOTs, elaborated in Section 3.4.1.} 


\begin{figure*}[htb]
    \centering
    \includegraphics[width=0.8
    \textwidth]{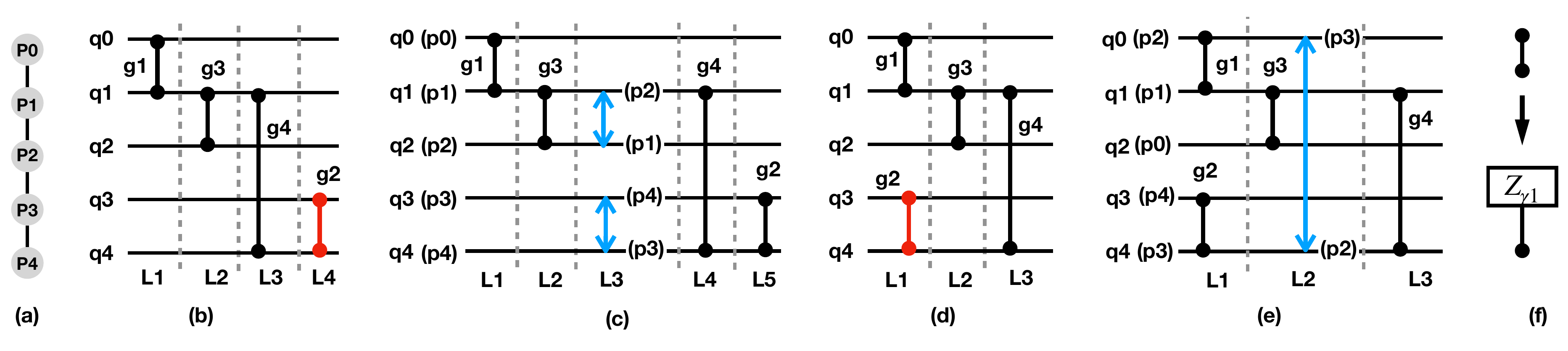}
    \caption{Different circuits result in different compilation results. (a) Hardware coupling; (b) One valid circuit corresponding to the input problem graph in Fig.\ \ref{fig:qaoaexample}; (c) One possible compiled circuit of (b); (d) Another valid circuit corresponding to the same problem graph; (e) Optimal compiled circuit of (d); (f) Control-Z gate abstracted. For the logical circuit (b), since its depth is already 4, its compiled circuit must have length $\geq$ 4. For the logical circuit (d), its optimal compiled circuit (e) does not increase its original depth 3. Therefore, the logical circuit (d) is definitely better than the logical circuit (b).   
 }
    \label{fig:qaoaexample_abs}
\end{figure*}

\section{Background} 
\label{sec:background}


\subsection{Introduction to QAOA}
The QAOA circuit is a multilevel parameterized quantum-classical hybrid circuit. It has $p$ optimization levels with two  parameters in each level. At each level it runs a circuit with the same structure, with the same set of gates except the rotation angles are changed. Each level is associated with a pair of parameters: $\gamma$ and $\beta$ specifying two different rotation angles for the controlled phase gates and the $RX$ gates. Fig.\ \ref{fig:qaoaexample} shows the circuit for QAOA-Maxcut for one level. The values of $\gamma$ and $\alpha$ at level $i$ are determined by machine learning on classical computers, typically stochastic gradient descent, and based on the quantum execution of previous $i-1$ levels. It is shown that QAOA has $78\%$ chance of obtaining an optimal solution in the first level \cite{QAOA:farhi2014quantum}.


\subsection{Compilation of QAOA}

\vspace{-5pt}
\subsubsection{Problem-graph induced circuit} Unlike other quantum applications, the structure of a QAOA circuit is not known until an input-dependent problem graph is specified. For instance the QAOA-Maxcut circuit is dependent on the input graph. In a QAOA-Maxcut circuit, a qubit corresponds to one vertex in the input problem graph. Each CPHASE gate corresponds to an edge in the problem graph, as shown in Fig.\ \ref{fig:qaoaexample}, the edges in (a) correspond to the two-qubit gates in (b) and (c).

What distinguishes QAOA circuit from other circuit is that the two-qubit operators commute. That is, they can flexibly permute, and the circuit outcome is the same. For instance, in Fig.\ \ref{fig:qaoaexample}, both (b) and (c) are valid instances of the QAOA circuit for the input problem graph in (a). The commutativity feature exists in 2-local Hamiltonian simulation circuits too. Such flexibility can be exploited to optimize compilation, for instance, reduce the gate count and depth of the compiled circuit, as described in Section \ref{subsubsec:compgoal}. 

\vspace{-5pt}
\subsubsection{Compilation objective}
\label{subsubsec:compgoal}

Near-term quantum computer's native gate set consists of single-qubit and two-qubit gates. A two-qubit gate applies to two connected physical qubits. Due to the instability of qubits and resonance between qubits and coupling edges, the connectivity between qubits on a quantum chip is limited. To tackle this problem, the locations of two logical qubits in a 2-qubit gate need to be remapped via SWAP gates until they reside on two connected physical qubits. A SWAP gate occurs between two connected physical qubits. Examples are shown in Fig.\ \ref{fig:qaoaexample_abs}, where SWAP gates are inserted to make two quantum circuits executable. The goal of compilation is to yield better gate count, depth, and fidelity.

of the compiled circuit.

The flexibility in gate ordering in QAOA circuit introduces new opportunities for optimization at compile-time. Two resulting circuits from the same problem graph can lead to different optimal compiled circuits. We show an example in Fig.\ \ref{fig:qaoaexample_abs} where one induced circuit results in no depth change (3) even when a SWAP gate is inserted, while the other circuit itself already incurs a larger depth (4) than the first one, even without adding any SWAPs. 


\subsection{2-local Hamiltonian Simulation}

Hamiltonian simulation is another important application of today's universal quantum computers. Hamiltonian simulation if performed on classically computers is an intractable problem, as its complexity grows exponentially with the problem size. Richard Feyman \cite{feynman:IJTP82}  is the first to propose to use quantum computers to efficiently solve the problem. 2-local Hamiltonian is a  type of Hamiltonian that arise in many physical systems \cite{lao+:arxiv21twoqan}. It has the following form:

\begin{equation}
    \begin{split}
        H = \sum_{uv \in E}{H_{uv}} + \sum_{k \in V}H_k
    \end{split}
\end{equation}

The interaction between the qubits in the 2-local Hamiltonian simulation can be represented by a graph $G = \{V, E\}$ where $H_{uv}$ is a two-qubit Hamiltonian and $u-v$ represents an edge in E, and $H_k$ is a single-qubit Hamiltonian where $k$ represents a node in $V$. With a further trotterization step for Hamiltonian simulation, all 2-qubit terms can commute flexibly \cite{lao+:arxiv20}. And Trotterization \cite{trotter:jstor59, lloyd:science96, suzuki:jmp91} is a common approach for building a quantum circuit on universal quantum computer to asymptotically approximate the time evolution of a Hamiltonian simulation.

\section{Our Method}
\label{sec:basismethod}

\begin{figure}[htb]
    \centering
    \includegraphics[width=0.4
    \textwidth]{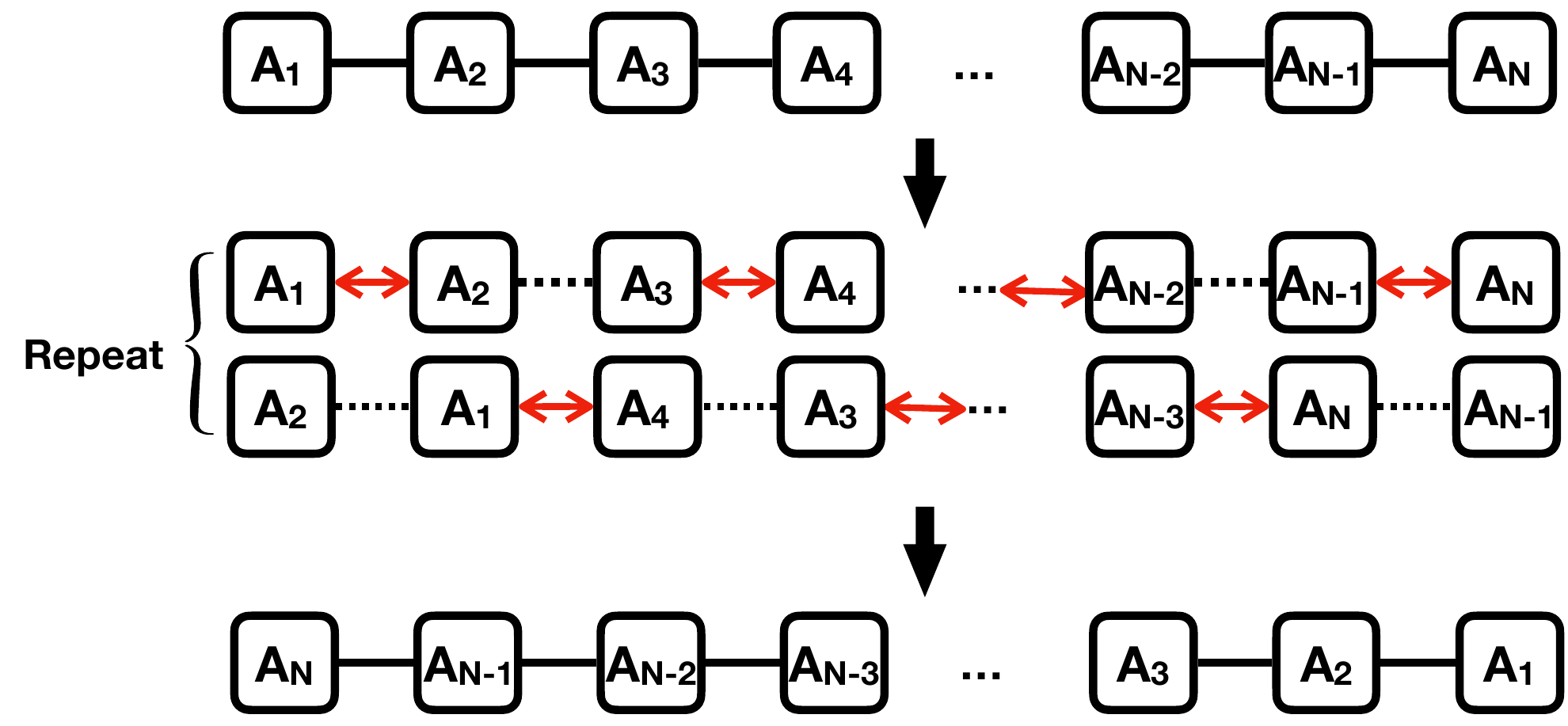}
    \caption{The line architecture solution.  }
    \label{fig:LAAPS_1D}
\end{figure}

\subsection{Solution for the Line Architecture}
Consider n nodes on a rope, labeled from $A_1$ to $A_N$. There exists a way to move each node around the rope, by swapping them with their neighbors, to form $\lceil N/2 \rceil$ different permutations such that each node is adjacent to every other node once and only once.

\begin{figure}
    \centering
    \includegraphics[width=0.4
    \textwidth]{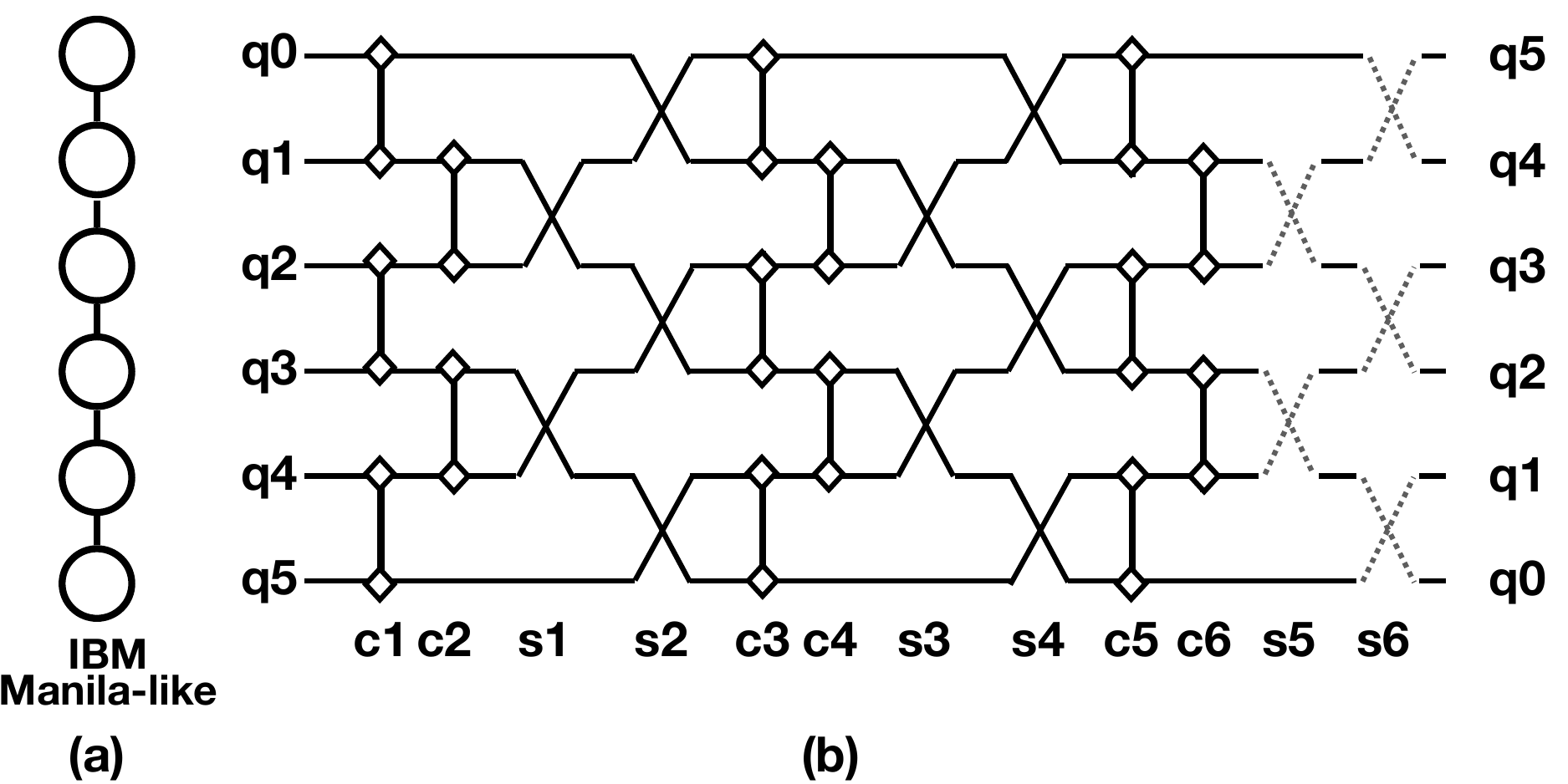}
    \caption{The solution on 6-qubit line architecture}
    \label{fig:laaps_on_manila}
\end{figure}

 The idea is shown in Fig. \ref{fig:LAAPS_1D}. Let the physical positions on the rope be labeled $P_i$, $i \in [1, N]$. First, nodes located in position $P_i$, where $i$ is odd, are swapped with their neighbor $P_{i+1}$. Then, nodes that are now located in $P_j$ where $j$ is even, are swapped with their neighbor $P_{j+1}$. A circuit corresponding to the idea is shown in Fig. \ref{fig:laaps_on_manila}. We purposely add an additional pair of SWAPs at the end which will result in the qubits being reversed in its final state. This last pair of swaps is not necessary to have all neighbors interact, only for reversing qubits.     Note that for odd-number qubits, in the last pair of SWAPs, we omit one for all-to-all connectivity. We do not discuss it due to space limit.

{The line architecture solution has been proposed in previous studies \cite{kivlichan+:prl18,ogorman+:arxiv19, weidenfeller+:arxiv22}. Overall it takes $N-1$ 
 SWAP layers (assuming each SWAP layer consists of parallel SWAPs) to achieve all-to-all connectivity and reversed qubit arrangement on the line. Certain representation lets one computation gate layer be followed by one SWAP layer \cite{kivlichan+:prl18, weidenfeller+:arxiv22}, and another representation lets two computation gate layers be followed by two SWAP layers \cite{ogorman+:arxiv19} but will cut one SWAP layer in the last pair. Ours is the latter. }


\vspace{-5pt}
\subsection{Solution for the 2xN Grid}
\label{subsec:2byngrid}

{The line architecture has been solved by previous studies as early as 2017 and 2019 \cite{kivlichan+:prl18, ogorman+:arxiv19}. However, the solution for a mutli-dimensional architecture has never been proposed until recently. We are the first to find a solution for the 2xN grid in 2021 \cite{our2021work} (in its Appendix). It is an important building block for the NxM grid solution. We believe  Weidenfeller \etal \cite{weidenfeller+:arxiv22} has a similar building block to this. It is also a building block for  Google Sycamore and the hexagon architecture whic we are the first to solve.} 

\begin{figure}[htb]
    \centering
    \includegraphics[width=0.4
    \textwidth]{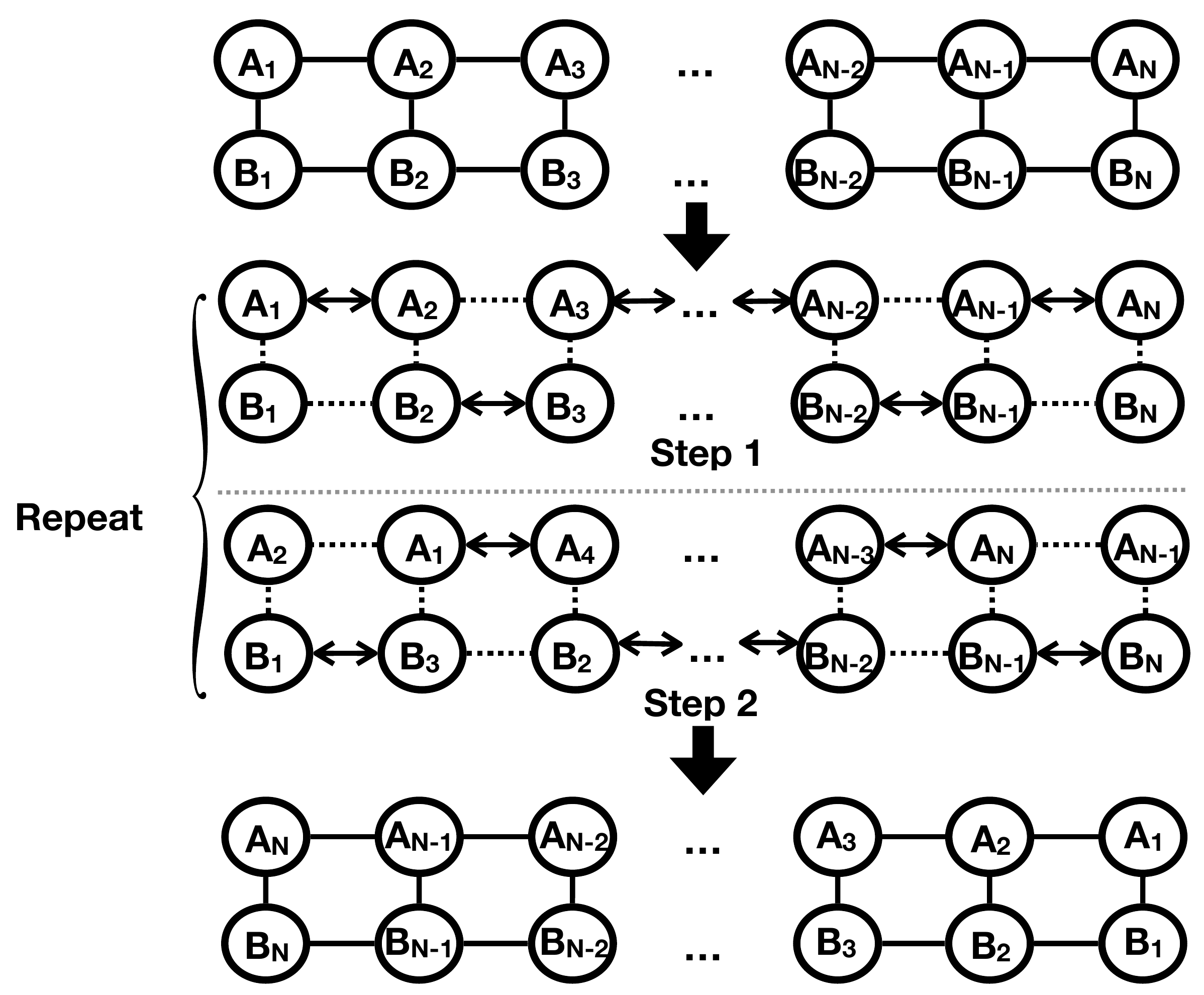}
    \caption{Solution for 2xN architecture.     Repeat $\lceil N/2 \rceil$ times to reverse the nodes on top row and bottom row respectively and ensure all-to-all interaction. }
    \label{fig:abstracted_2xN}
\end{figure}

The idea is simple, as shown in Fig. \ref{fig:abstracted_2xN}. Note that each row performs a swap pattern that mirrors that for the line architecture, but the swaps alternate between adjacent units. The reason for the alternating swap pattern is as follows: imagine two parallel strings of nodes such that each node has a partner node on the other string (vertically). If the nodes on both strings move in the same direction, then the nodes will always have the same partner. This movement pattern enable both inter-row and intra-row all-to-all interaction. Interestingly, all intra-row interactions happen simultaneously (one parallel layer) at each step, and all inter-row interactions happen simultaneously, which means they can be separated if necessary. 


\subsection{Handling Multi-dimensional Architecture}

To handle multi-dimensional architecture, our idea is to break down an $N$-dimensional architecture into multiple $(N-1)$-dimensional units. Assuming there are $M$ units, by looking at the connectivity between these units, it is at least a line. If it has more connectivity than a line, that's okay too. 


If a few condition can be met,  we can handle the N-dimensional architecture, by just considering every unit as if it is a ``qubit" on a line, every unit-exchange as if it is a ``SWAP" in a line, and every inter-unit interaction as if it is a ``two-qubit gate" between two qubits. Note that the sub-problem of the N-1 dimensional architecture can be handled in the same way. 
The Maaps algorithm is shown in Algorithm \ref{alg:MAAPS}.


\begin{algorithm}
\caption{Maaps on multi-dimensional architecture}
\label{alg:MAAPS}
\begin{algorithmic}[1] 
\Procedure{Maaps}{$P$}:
    \State Divide P into M units: $P_0$, $P_1$ ... , $P_{M-1}$
    \If{$\mid P_i\mid $ != 1}
        \State MAAPS($P_i$);  $~~$           i=0, 1, ... M
        \For{s=0; s+=1; s<M/2}
        \State IE($P_i$, $P_{i+1}$); $~~$     i=0, 2, ... M
        \State IE($P_j$, $P_{j+1}$);  $~~$    j=1, 3, ... M
        \State UE($P_j$, $P_{j+1}$);  $~~$    j=1, 3, ... M
        \State UE($P_i$, $P_{i+1}$); $~~$    i=0, 2, ... M
        \EndFor
    \Else
        \State P is a line now. Handle it as a line architecture. 
    \EndIf
\EndProcedure
\end{algorithmic}
\end{algorithm}


We formalize the four conditions that must be satisfied to handle any multi-dimensional at each recursive step:
  
\begin{itemize}
  \item Define "units" such that the connectivity of the units forms a line.
  \item Define intra-unit (IA) operations. That is, all-to-all interaction within a unit, in linear time.
  \item Define inter-unit (IE) operations: Having all-to-all interaction between two neighboring units, in linear time. 
  \item Define unit exchange (UE) operations: Exchanging the locations of two neighboring units in linear time. 
\end{itemize}



 
With the four above conditions, Maaps can ensure linear time all-to-all interaction.  It can be proved rigorously. We sketch the proof. Assuming the $N-1$ dimensional sub-space size is X, and the number of units is $M$. The total size is $M*X$ for the N-dimensional space. Assuming the $(N-1)$-dimensional sub-problem has linear solution O(X). Since the inter-unit interaction happens $2M+O(1)$ times and is linear, and the intra-unit interaction is linear $O(X)$, the complexity is then $O(M*X)$  linear to the size of the entire N-dimensional space.

\begin{figure}[htb]
    \centering
    \includegraphics[width=0.3
    \textwidth]{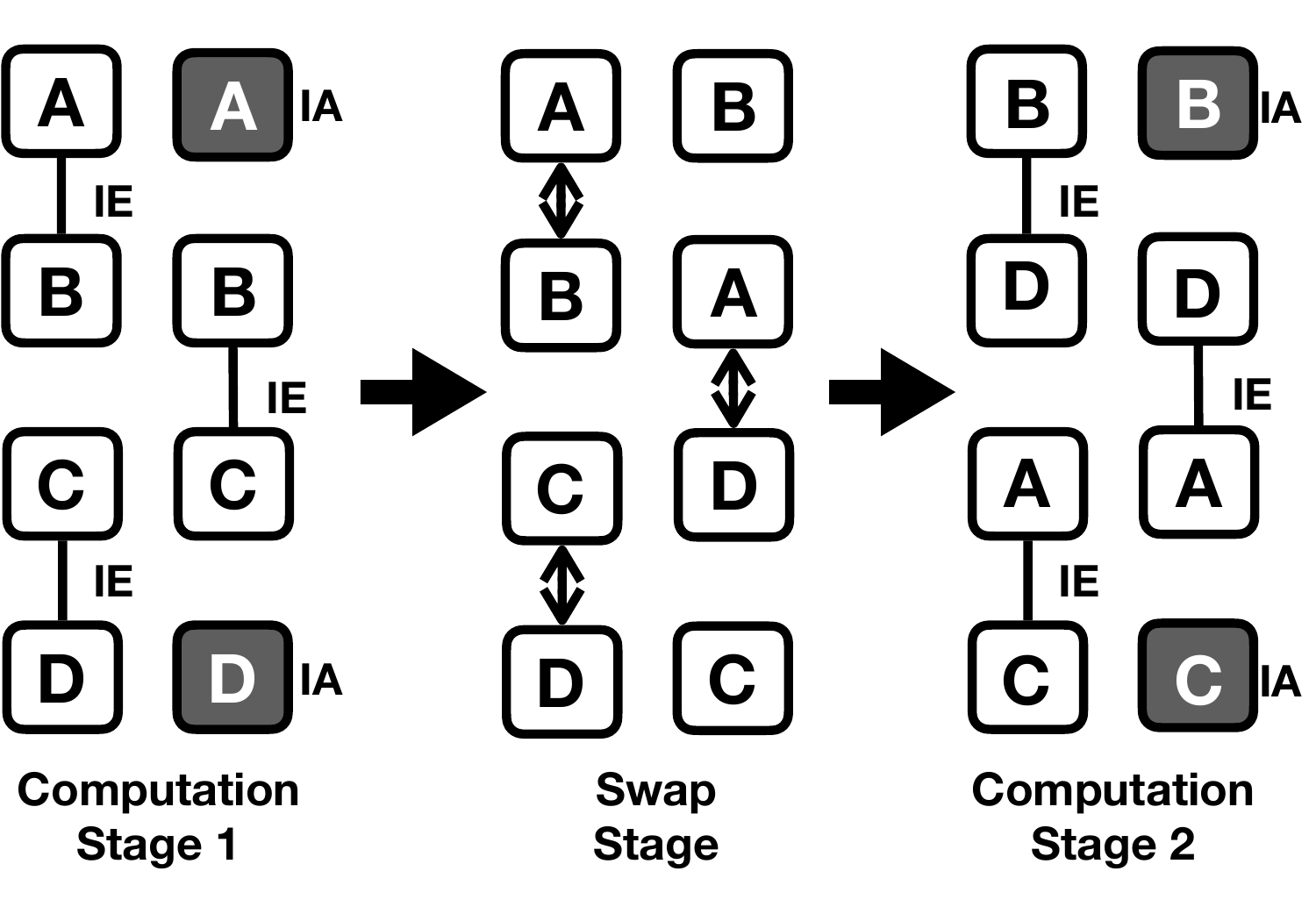}
    \caption{Row-wise movement and interaction for 2D lattice. Each node represents a row. Assuming four rows A, B, C, and D. A and D in computation stage 1 are idle and hence they can perform IA operations. Similarly for B and D in computation stage 2.  } 

    \label{fig:intra_inter}
\end{figure}

\subsection{Google Sycamore, 2D Lattice, {and Hexagon}}

 We demonstrate how Maaps applies to 3 important architectures: 2D lattice, Sycamore, and hexagon. We start with the 2D lattice example as it illustrates the application of Maaps most clearly.

\subsubsection{2D Lattice}
\label{subsec:2dlattice}
\noindent \textbf{\underline{Defining A Unit}}:
For 2D lattice, we define a unit as a row. 

\noindent \textbf{\underline{Unit Exchange}}: Unit exchange is simply applying vertical pairwise swaps between two rows. One parallel level of such swaps exchanges the locations of the entire two rows.

\noindent 
\textbf{\underline{Intra-row (IA) Operation}}:
Intra-row all-to-all interaction can follow that for the linear architecture, which is trivial.


\begin{algorithm}
\caption{Inter-unit interaction for two rows in 2D lattice}
\label{alg:MAAPS-Lattice}
\begin{algorithmic}[1] 
\Procedure{$IE_{Lattice}$}{A, B}:
    \State r = 0
    \State s = sizeof(A)
        \For{i=0; i+=2; i<s}
          
            \State $G_{interact}$($A_{j}$,$B_{j}$);  j=0, 1, ... s
            \State SWAP($A_{j}$, $A_{j+1}$);     j=r, r+2, ..., s
            \State SWAP($B_{j}$, $B_{j+1}$);     j=1-r, 3-r, ..., s
            \State r = (r+1) mod 2
    \EndFor
\EndProcedure
\end{algorithmic}
\end{algorithm}

\noindent \textbf{\underline{Inter-unit Interaction}}: The goal of the IE interaction in 2D lattice is to ensure that every node in unit (row) $A$ interacts with every node in its neighboring unit (row) $B$. We can apply our solution for 2xN grid in Section \ref{subsec:2byngrid}. Recall that intra-row and inter-row operations can be decoupled into separate layers and either runs  within minimal number of layers guaranteed by our 2xN solution, we only keep the inter-row operations.We present a concrete example in Fig. \ref{fig:2x3_example}, which is derived from our previous example \cite{our2021work}.   We outline the algorithm for IE interactions on a 2D lattice in Algorithm \ref{alg:MAAPS-Lattice}.  
 
 {Just naively applying the general Maaps method may not provide the best solution. We notice two optimization opportunities that happen to work for 2D grid case. The work by Weidenfeller \etal \cite{weidenfeller+:arxiv22} also includes one of them, but not both. Hence, ours is slightly better, which reduces 25\% depth. We describe them below. }

\begin{figure}
    \centering
    \includegraphics[width=0.45\textwidth]{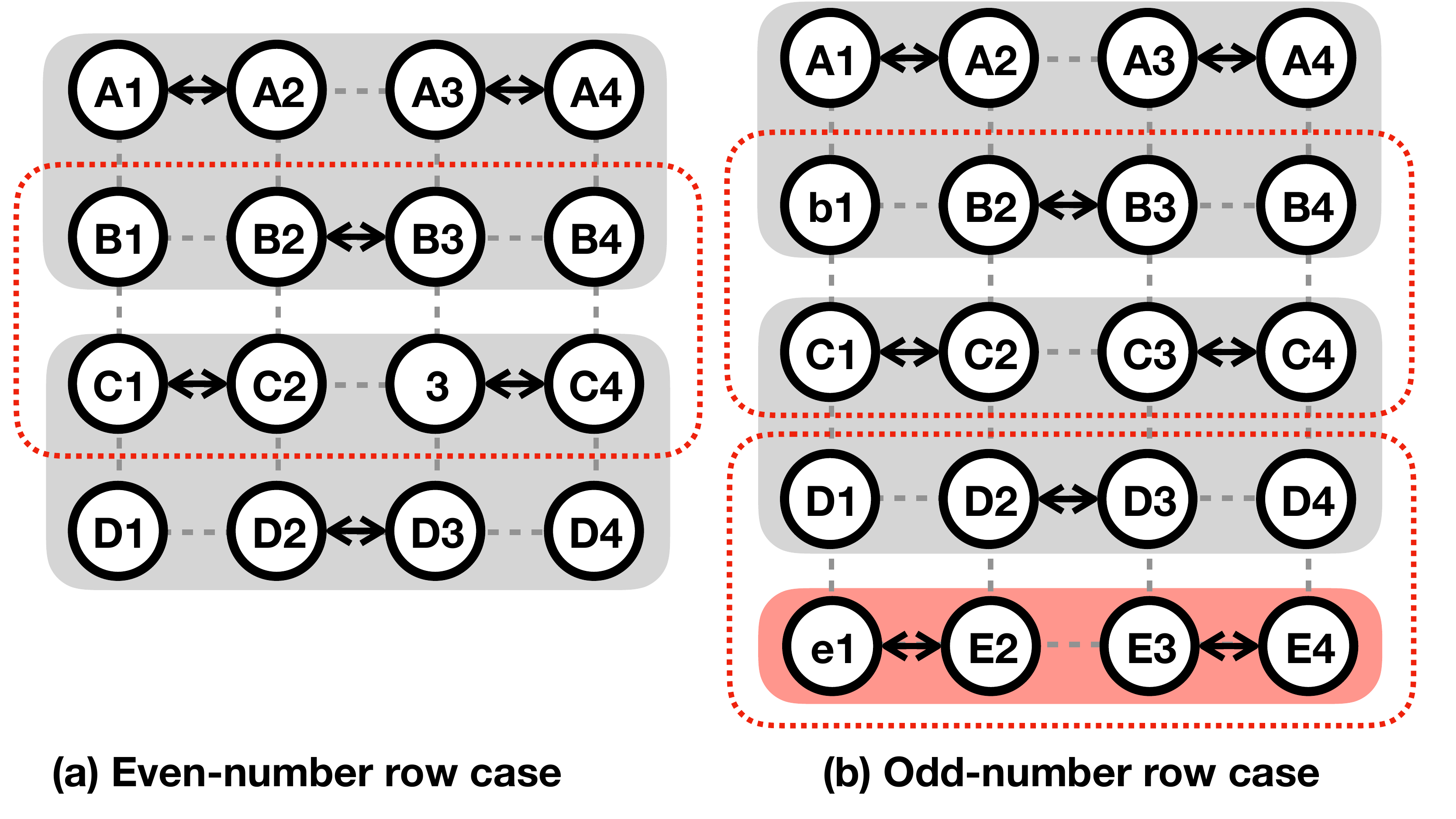}
    \caption{Grid Optimization I: (a) Odd-even row SWAP happens to help even-odd row interaction, (b) One additional SWAP layer added in addition to odd-even row SWAP.}
    
    \label{fig:gridoptI}
\end{figure}
 
 \noindent\paragraph{{{{Further Optimization I -- Sharing SWAP Layer.}}}} In our representation of the line solution, after each pair of SWAP layers, we have two computation layers, one on odd-even units, and the other on even-odd units.  An example is shown in Fig. \ref{fig:big2Dstep}, where in steps (e) and (f) \{A-B, C-D\} interaction is odd-even unit IE, and B-C is even-odd unit IE.

In our general Maaps method, even-odd unit IE needs to be completed before odd-even IE operations, or the other way around, but we can mix them for the grid case. 

 We notice that the SWAP layers performed for the odd-even row pair can prepare for the layout change needed for the even-odd pair. In a 2xN operation, the top-row performs odd SWAPs (where the left index of a SWAP is an odd number, assuming the index starts from 1), and the bottom-row performs even SWAPs, or the other way around, as shown in Fig. \ref{fig:gridoptI} (a) for rows A and B in grey boxes, and for rows C and D. It happens that it also creates a SWAP pattern that enables all-to-all connectivity  for the even-odd row IE, as highlighted in red dashed box for rows B and C in Fig. \ref{fig:gridoptI} (a). Hence we can use just one SWAP layer for both odd-even row and even-odd row IE operation.

{A subtle point to note is that when the number of rows is odd. Just preparing the SWAPs in the odd-even pairs of rows would not be enough, since the last row will be idle and no layout change is performed, as shown in Fig. \ref{fig:gridoptI} (b). But is necessary for the last even-odd row pair's IE. Therefore, we can add one more SWAP layer at the last row in parallel, as shown in Fig. \ref{fig:gridoptI} (b) for row E. We can do this for 2D grid, since the last row's nodes are connected via a line.  But this does not hold for Sycamore and hexagon.}

\begin{figure}[htb]
    \centering
    \includegraphics[width=0.45
    \textwidth]{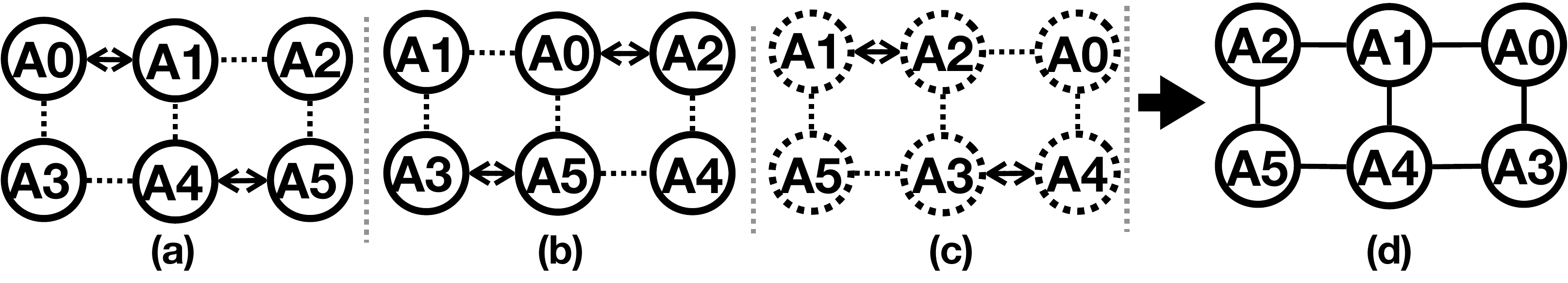}
    \caption{An example of swap pattern for IE of 2D lattice. On the two rows, we have alternating SWAPs. (a), (b), and (c) are three swap steps. It can be seen that based on the first two SWAP layers, $A_0$ will have $A_3$, $A_4$, and $A_5$ as its partner, which covers $A_0$'s interaction with all elements in the second row. The same applies to other elements $A_i$. The last SWAP at (c) is not necessary, we just use it to show the symmetry. {This is derived from the example in our work in 2021 December \cite{our2021work} in 2021. }
 }
    \label{fig:2x3_example}
\end{figure}

\noindent {\paragraph{Further Optimization II: Interleaving IE and IA.}}
The second optimization we apply is to overlap IA and IE operations. So far we haven't talked about when to handle intra-unit (IA) case. It is possible that one can run IA for all rows at once since within each row it forms a linear line. In fact, we can even save this step by overlapping IA operations with IE without any additional cycles.

{ During the step of  pair-wise IE: even-odd row IE, and odd-even row IE, there is always an unit or two that is idle. If the total number of rows is even, then during the even-odd pair of IEs, the first row and the last row, are idle. If the total number of rows is odd, then for the odd-even pair of IE, the last row is idle, and for the even-odd pair of IE, the first row is idle. For two consecutive even-odd and odd-even IE operations, there are always two rows that are idle. We can perform intra-unit (IA) operation for these idle rows at the same time as the IE operations. It happens that an IE cycle depth is the same as an IA cycle depth in our prior work \cite{our2021work} using the 2xN solution. }
 
{Note that this will cover all rows, since according to the unit movement pattern in the line proved by previous work \cite{weidenfeller+:arxiv22} and our non-refereed publication \cite{our2021work}, each node is moving towards a direction until it hits the border, then it changes the direction.  Each node stays at the first or last location for exactly once. An example showing that it will cover all units in top row or bottom row is shown in Fig. \ref{fig:intra_inter}. }


\begin{figure*}[htb]
    \centering
    \includegraphics[width=0.85
    \textwidth]{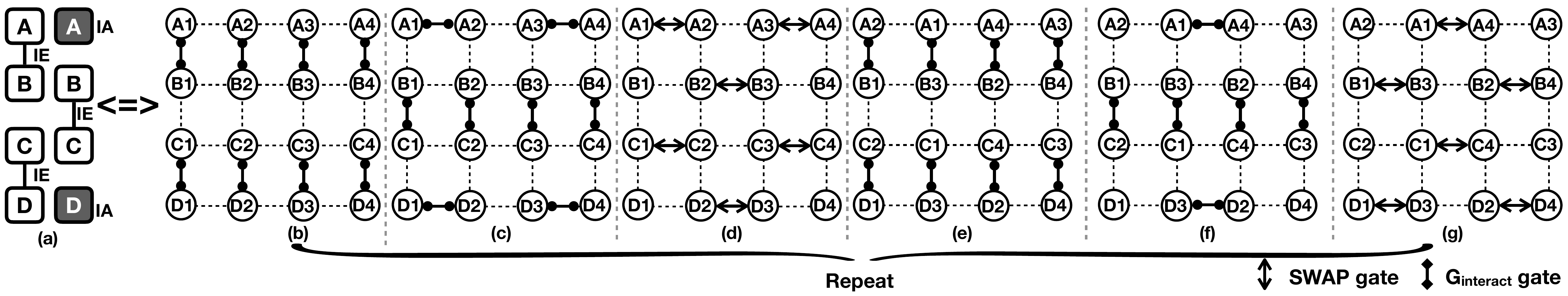}
    \caption{A detailed step  for interleaving intra- and inter-unit operations, corresponding to computation stage 1 in Fig.\ \ref{fig:intra_inter}.  
 }
    \label{fig:big2Dstep}
\end{figure*}

{This optimization is not present in the concurrent work by Weidenfeller \etal \cite{weidenfeller+:arxiv22}. In total, they have one SWAP layer followed by three computation layers (two IE and one IA), while ours has one SWAP layer followed by two computation layers (one IE and one IE/IA). We can reduce the depth by 25\%. Here we do not decompose SWAP into 3 CNOT or cancel CNOTs.}\\ 

\noindent \textbf{\underline{Putting it all together}}: 

{With the two optimizations, we can reduce the depth significantly, as shown in a full example in Fig. \ref{fig:big2Dstep}. 


To calculate the depth, we assume it is a square lattice, both the row and column numbers are N, without loss of generality. 
It takes 3/2 cycles for each IE$_{Lattice}$(A, B)'s iteration on average since the two SWAP steps (line 6 and 7).  This will take a total of 3N-1 cycles (no need to swap for last iteration) since it takes N iterations for each node in a row to be neighbors with every other node in another row. These 3N-1 cycles repeat N/2 times since there are N/2 unit configurations that must be reached for all units to be neighbors with each other once. 

There is a total of N-2 unit exchanges since there are N/2 configurations that need to be reached, but it takes 2 swap cycles to reach each configurations. No swaps are needed after the final configuration, hence the N-2 total swaps. Thus, the total run time is (3N-1)(N/2) + (N-2) cycles.

{For method by Weidenfeller \etal \cite{weidenfeller+:arxiv22}, it does not overlap IE with IA. Rather it runs IE and IA separately after each set of horizontal SWAPs (two IEs and one IA, see Fig. 17 in their paper).  Hence their depth is $2N^2+O(N)$ in total. We reduce 25\% depth compared with theirs.}


\begin{figure}
    \centering
    \includegraphics[width=0.4\textwidth]{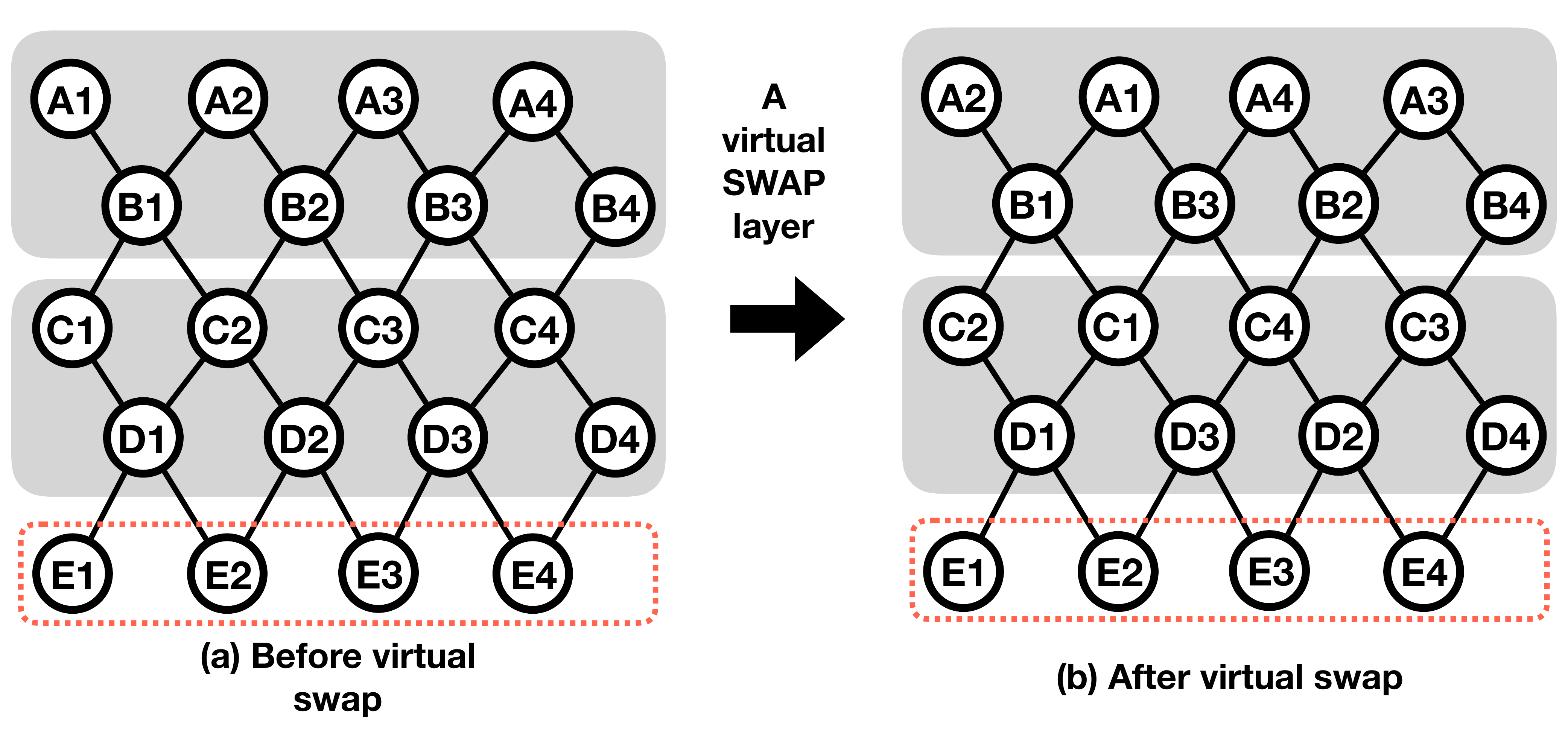}
    \caption{Odd-number row case for Sycamore}
    \label{fig:sycamore_odd}
\end{figure}

\begin{figure}[htb]
    \centering
    \includegraphics[width=0.40
    \textwidth]{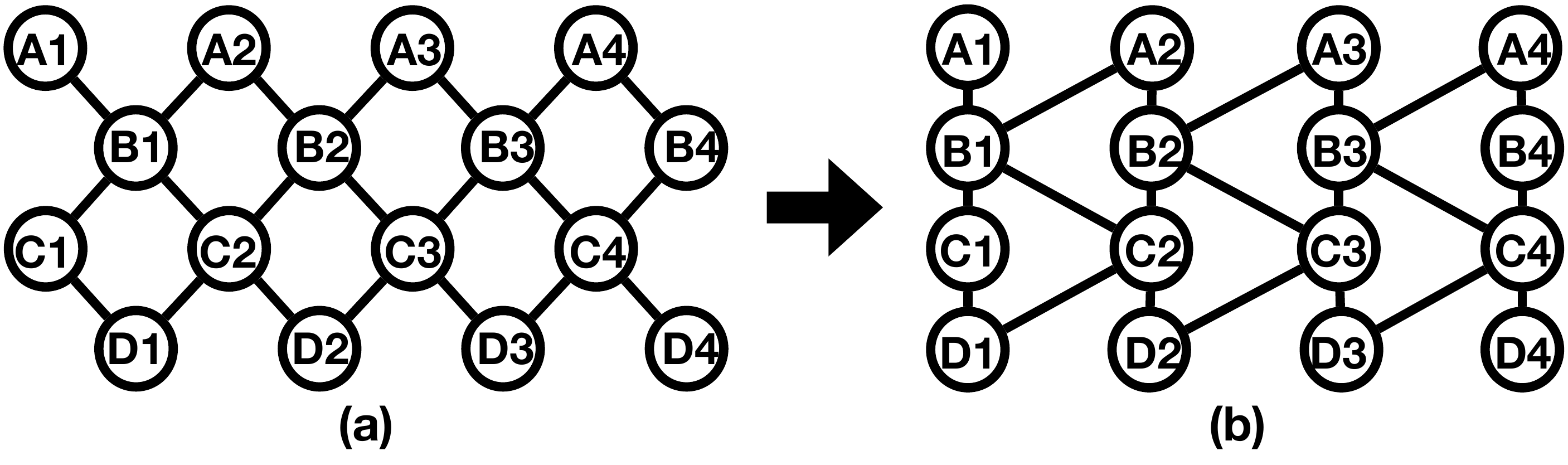}
    \caption{Sycamore illustrated in a different way.}
    \label{fig:2draged_sycamore}
\end{figure}

\begin{figure*}[htb]
    \centering
    \includegraphics[width=1
    \textwidth]{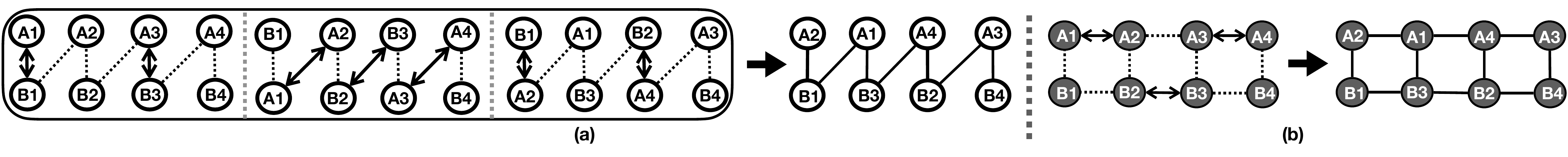}
    \caption{(a)Sycamore takes 3 swap cycles (one virtual SWAP layer to mimic the 2xN lattice inter-row interaction, (b) Corresponding 2xN lattice operation requires just 1 swap cycle}
    \label{fig:2Dswap_To_sycamore}
\end{figure*}

\subsubsection{Google Sycamore}

\begin{algorithm}
\caption{A Mixed IE and IA Step for MxN Sycamore}
\label{alg:MAAPS-Sycamore}
\begin{algorithmic}[1] 
\Procedure{IEIA$_{Sycamore}$}{$P$}:
    \State r = 1
    \For{each unit $P_i$, i is even (starts from 2), in P}
            \State \textbf{[1]} $G_{interact}$($P_{i, j}$,  $P_{i+1, j}$);  $~~$ j=1, 2, ... M
            \State \textbf{[2]}
            $G_{interact}$($P_{i, j}$, $P_{i-1, j}$);  $~~$ j= 2, 4, ... M 
            \State \textbf{[3]} SWAP($P_{i, j}$, $P_{i-1, j}$); $~~$   j=r, r+2, ... M
            \State \textbf{[4]} $G_{interact}$($P_{i,  j}$, $P_{i-1, j+1}$); $~~$ j=1, 2, ... M-1
            \State \textbf{[5]} SWAP($P_{i, j}$, $P_{i-1, j+1}$);  $~~$  j=1, 3, ... M
            \State \textbf{[6]} $G_{interact}$($P_{i, j}$, $P_{i+1, j}$); $~~$    j= r, r+2, ... M
            \State r = (r+1) mod 2
    \EndFor
\EndProcedure
\end{algorithmic}
\end{algorithm}

 For illustration purpose, we redraw the  Google Sycamore case to look like a 2D lattice as illustrated in Fig.\ \ref{fig:2draged_sycamore}.  Note that this does not reduce the complexity of the problem.   \\
 
 \noindent \textbf{\underline{Defining A Unit:}} We define a unit still as a "row" in the redrawn Sycamore architecture in Fig.\ \ref{fig:2draged_sycamore}. For instance,  $A_1$, $A_2$, $A_3$, and $A_4$ form a unit.

  \noindent \textbf{\underline{Unit Exchange:}}
 Unit exchange then is trivial and only requires the vertical links.

 \noindent \textbf{\underline{\textbf{Inter-unit Interaction:}}} Recall that in the 2xN architecture, we perform odd SWAPs and even SWAPs on two rows repeatedly. This ensures each node on the top row interacts with each node in the bottom row  through vertical links, as shown in Fig.\ \ref{fig:abstracted_2xN}. 
 
 We now implement Sycamore's IE by mimicking the this behavior of 2xN.  We use the diagonal/vertical links between rows to achieve this goal but with more cycles. All together we use 3 SWAP layers to achieve this goal, which we refer to as a ``virtual horizontal SWAP" layer. It takes 2 vertical SWAPs, and one diagonal SWAP. Our idea is shown as an example in Fig.\ \ref{fig:2Dswap_To_sycamore}, it is for mimicking the top-row \emph{odd} SWAP and bottom-row  \emph{even} SWAP case in grid. A similar set of 3 SWAPs can be achieved for mimicking the top-row even SWAP and bottom-row odd SWAP case in grid. 
 
Now since all vertical links are maintained compared with the grid case, we will be able to perform inter-``row" computation after the ``virtual horizontal swap" layers. 

\noindent \textbf{\underline{Intra-unit Interaction:}} 



We notice that IA can be scheduled in between the actual SWAP layers of the ``virtual horizontal SWAP" layer.  Note that after the first SWAP layer, it connects the once separated qubits within a unit through the diagonal links. For instance, in Fig. \ref{fig:2Dswap_To_sycamore} $A1$ would now be connected with $A2$ and $A3$ with $A4$. Thus we can insert an IA operation amid the steps of an IE operation, as shown in Fig.\ \ref{fig:sycamore_big_step} (e).

\begin{figure*}[htb]
    \centering
    \includegraphics[width=0.85
    \textwidth]{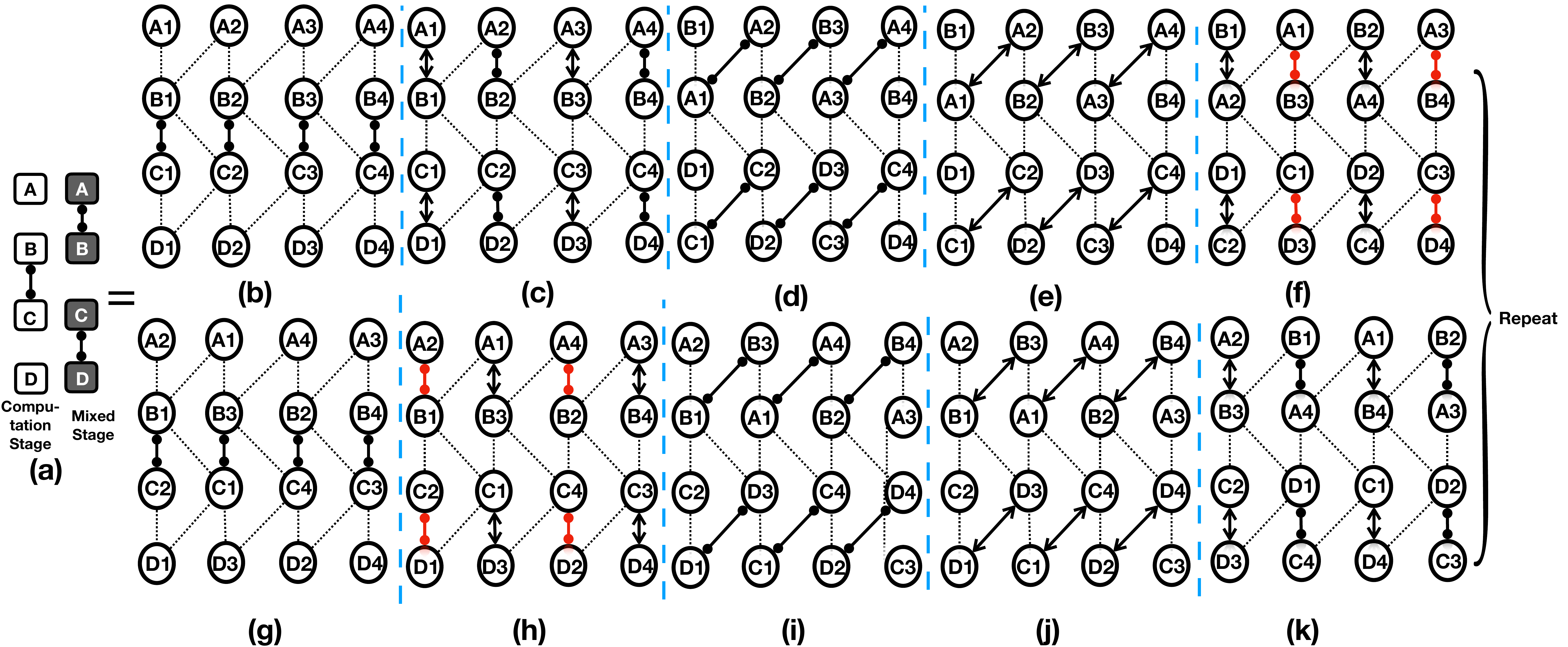}
    \caption{Basic Sycamore IA-IE Maaps pattern with interleaving. Repeat $N/2$ times.}
    \label{fig:sycamore_big_step}
\end{figure*}

\begin{figure}[htb]
    \centering
    \includegraphics[width=0.2
    \textwidth]{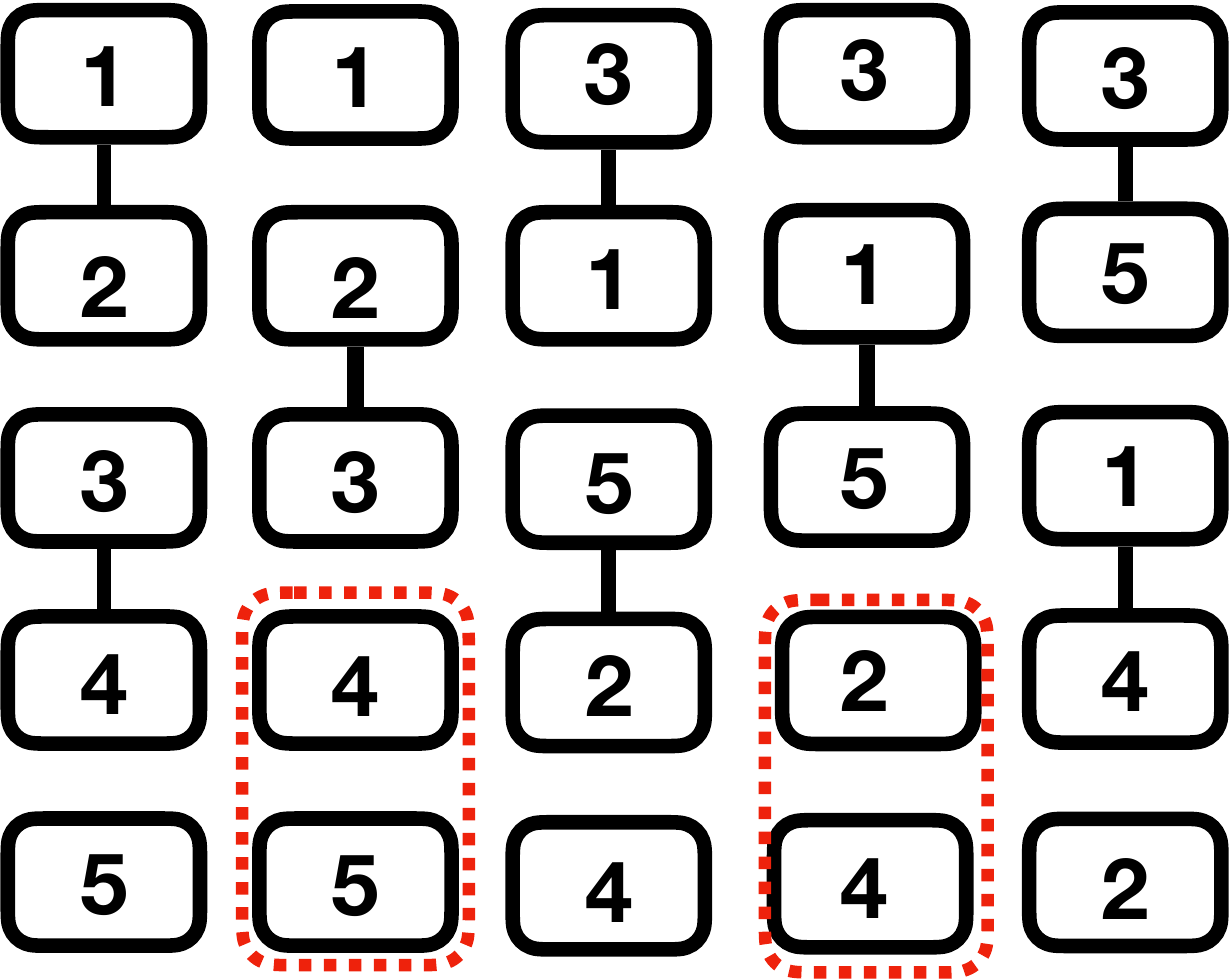}
    \caption{In odd-numbered row Sycamore, omitted IE operations for the bottom two physical rows after each pair of SWAP layers. Each node represents a unit. }
    \label{fig:odd_example}
\end{figure}


\noindent{\underline{Optimization for Overlapping IE and Virtual SWAP:}} Since during two SWAP layers within a virtual SWAP layer, half of the links remain idle, it allows us to schedule half the IE operations on these idle links. An example is shown in Fig.\ \ref{fig:sycamore_big_step} (c), (f), (h), and (k). It is as if a vertical gate interaction between two rows are split into two steps, shown as the red links (gates) in Fig. \ref{fig:sycamore_big_step} (f) and (h). We omit the discussion of the head and tail here due to space limit.

\noindent{\textbf{\underline{Putting everything together}}}: We describe the interleaved IE and IA operations in Algorithm \ref{alg:MAAPS-Sycamore} and illustrate it with an example in Fig.\ \ref{fig:sycamore_big_step}. Note that both illustrate IE operations for all rows and columns in the Sycamore lattice, not just a pair of rows.  In total it takes 5 cycles for such a big iteration (excluding the last iteration where certain swaps can be omitted) although we have drawn 8 of them. 

{
\paragraph{Discussion of the Odd-number Row Case}
One can see that we performed a similar optimization to Optimization I in the grid case in Section \ref{subsec:2dlattice}. We let one virtual SWAP layer be shared by two consecutive odd-even row and even-odd row IEs. For instance, in Fig. \ref{fig:sycamore_big_step}, after a virtual SWAP layer (c), (e), and (f), the computation on (g) as well as the red links on (f) and (h) cover even-odd and odd-even row IEs respectively.  This would still work for the even-number row case for Sycamore, but not for the odd-number row case. The last row in the odd-number row case does not have direct links between qubits within it, not like the last row in the grid.  Moreover, since all other rows are busy now, it cannot use the  diagonal links.  Therefore we cannot satisfy the layout change needed for the even-odd row pair (the last and second-to-last rows). 

In this case, we can fall back to the generic Maaps method, by not sharing the SWAP layer for two consecutive odd-even and even-odd row IE. Asymptotically, we increase the depth by a factor of $6/5$. But we do not have to do so. Our idea is to simply neglect the interaction between the last physical row and the second to last physical row. Then we perform the missing inter-row IEs in a separate pass. The idea of omitting is shown as an example in Fig. \ref{fig:odd_example}. 

It happens that the omitted IEs  are between consecutive even-numbered rows, and between the second-to-last row and the last row. We sketch the proof here. Since even-numbered units move down the line until it hits the border and stops there for one SWAP cycle before it changes direction \cite{weidenfeller+:arxiv22, our2021work}, each even numbered row will hit the bottom border once. Also since originally the distance between two even-numbered units is 2 and they move at the same speed before hitting the border, the last two physical rows will contain two even-numbered rows except one case. During the very first step, the IE between the second-to-last row (the largest even-numbered row) and the last row is also missing. 

With this property, we can easily break down the missing IEs into two parallel layers. The first layer consists of pairs of rows 2-4, 6-8, 10-12, and so on. The second layer consists of the rest of the pairs, 4-6, 8-10, and so on. Since the distance between each pair is 2, we can apply concurrent SWAPs to move one next to the other. Then we revert the SWAPs to restore the natural number ordering, then we do the same for another parallel layer. 
In total it takes 6 layers including SWAP, computation and reverting, but it applies the factor in front of $N$ which is the square root of the total qubit number $N^2$, as each IE takes $N-1$ depth.  

}

\noindent \textbf{\underline{Runtime}}: 
The IA loop is 5 cycles for an NxN  Sycamore, the set of cycles will take 5N-1 cycles. Thus, the total run time on any Google Sycamore architecture is at most (5N-1)(N/2) + (N-2) cycles. Compared with our grid solution, it is 5/3 factor slower. Compared with the line architecture, it is 5/4 factor slower. Note that it is not possible to draw a line to connect all nodes in Sycamore. Hence we cannot directly apply the line architecture solution here.

\vspace{-5pt}

\begin{figure}[htb]
    \centering
    \includegraphics[width=0.35
    \textwidth]{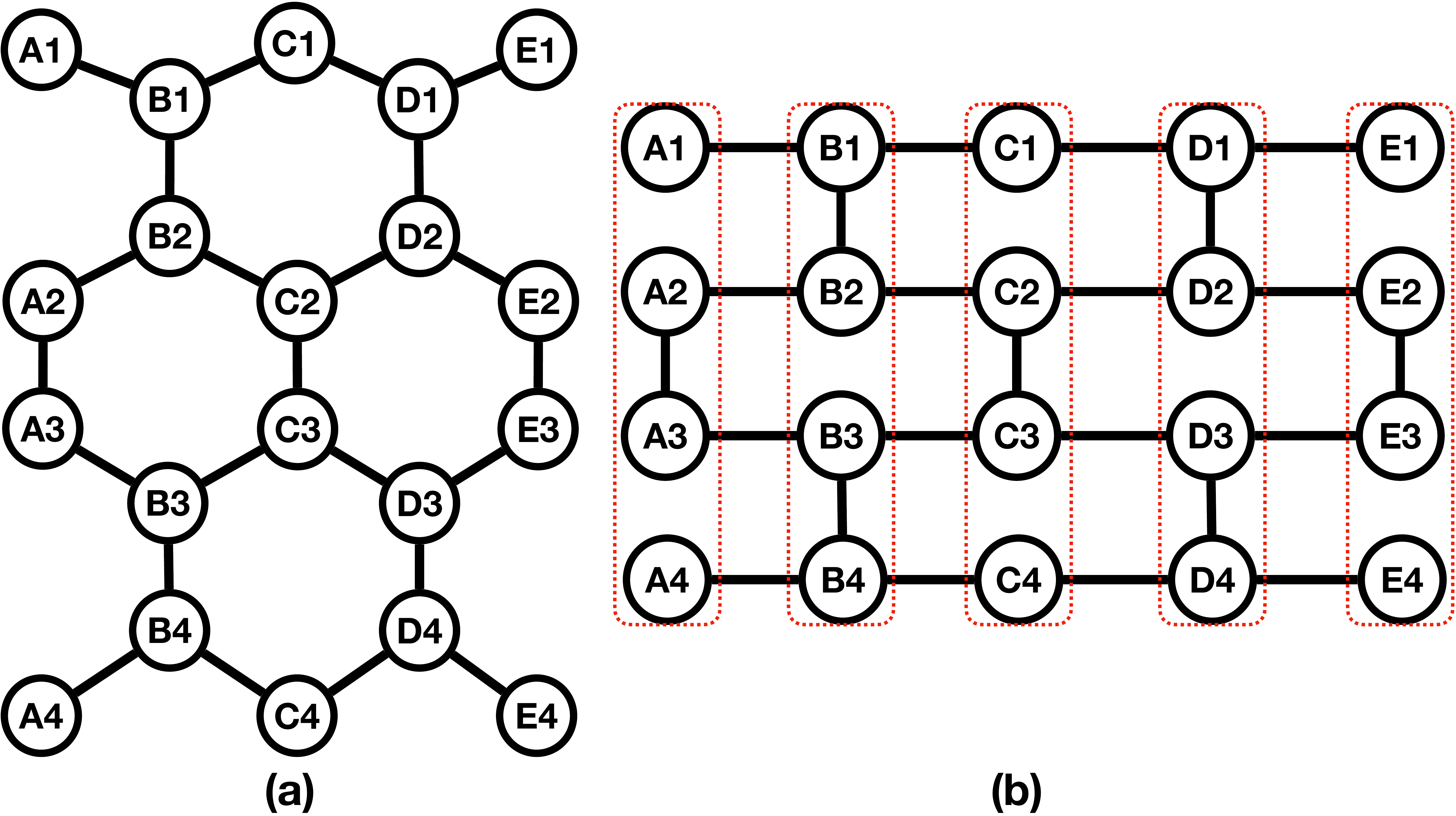}
    \caption{Hexagon coupling graph.}
    \label{fig:hex_coupling}
\end{figure}

\begin{figure}[htb]
    \centering
    \includegraphics[width=0.45
    \textwidth]{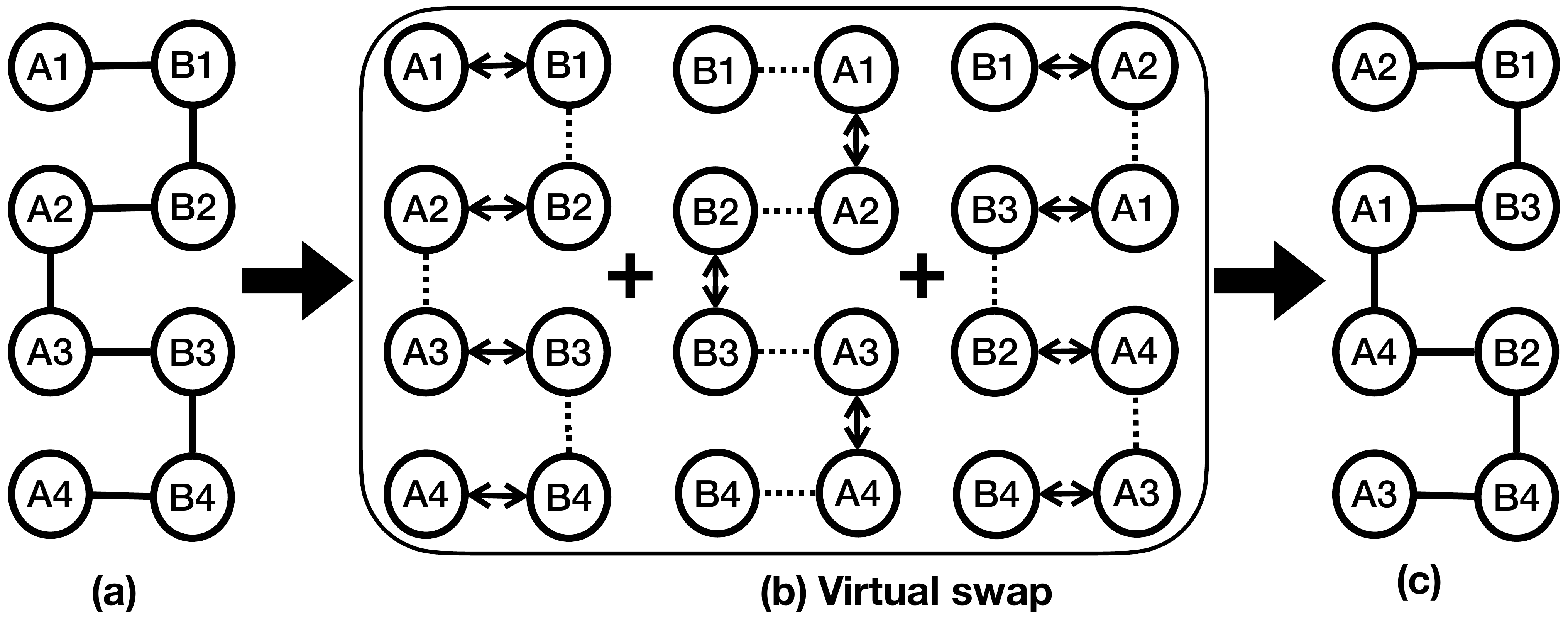}
    \caption{A virtual SWAP layer for the hexagon structure.}
    \label{fig:hex_swap}
\end{figure}

{
\subsubsection{The Hexagon Architecture}

Here we present an imaginary architecture -- the connected hexagon architecture in Fig. \ref{fig:hex_coupling} (a). Each node except the boundary node has a degree of 3. If we flatten it out, it looks like the one in Fig. \ref{fig:hex_coupling} (b). Notice that there is often no line embedding that covers every node in the hexagon architecture, as shown in Fig. \ref{fig:hex_coupling} (b), so the line architecture solution cannot apply here.  We let each unit be a vertical column of nodes, as highlighted in Fig. \ref{fig:hex_coupling} (b). 

For unit exchange, it is simple. We use the direct horizontal one-to-one links. For inter-unit gate operation, we again mimick our 2xN architecture solution, as if we apply a virtual SWAP layer, as shown in Fig. \ref{fig:hex_swap}. This characterizes the connection between two consecutive units. Between odd-even units is the particular example in Fig. \ref{fig:hex_swap}. If you mirror it, it will work for between even-odd units. Therefore we solved the IE operations. For intra-unit operation, since every two units can form a line, it is solved. 

Last but not least, the hexagon architecture also has the odd-numbered unit problem as Google Sycamore. We solve it in the same ways as Google Sycamore.}

\subsection{Application on Non-Linear IBM Machines}
We can handle IBM heavy-hex architectures, by adapting the line solution.

Although there is no line embedding that covers all the nodes in these IBM architectures, we can still find a line embedding that covers the overwhelming majority of them. In Fig.\ \ref{fig:ibm_pattern}, we show an example of one such line embedding, with the gray nodes being off the line embedding. 

A very simple approach would be to run {two full sets of iterations for the line architecture in a row. It will ensure interaction between all nodes in the line, as well as the interaction between nodes on and off the line. As mentioned previously and proved in prior and our studies \cite{weidenfeller+:arxiv22, our2021work}, a node (whether it is even-numbered or odd numbered) moves towards a certain direction until it hits the border, once it hits the border, it stops for one step, and then changes the direction. In total it travels N steps with one step being stopped. It may not visit all physical locations, but if we let it continue and do another set of full iterations, it will cover each physical location, as shown in Fig. \ref{fig:ibm_2_iters}. That ensures a node on the line is neighbor to each off-line node.} Next we swap the nodes off the line onto the line and then run the line solution again. This guarantees to have all the nodes interact with each other in linear time. {The total depth is 6n+O(1) if the input graph is clique.} The idea is illustrated in Fig.  \ref{fig:ibm_pattern}. 





\begin{figure*}[htb]
    \centering
    \includegraphics[width=0.7
    \textwidth]{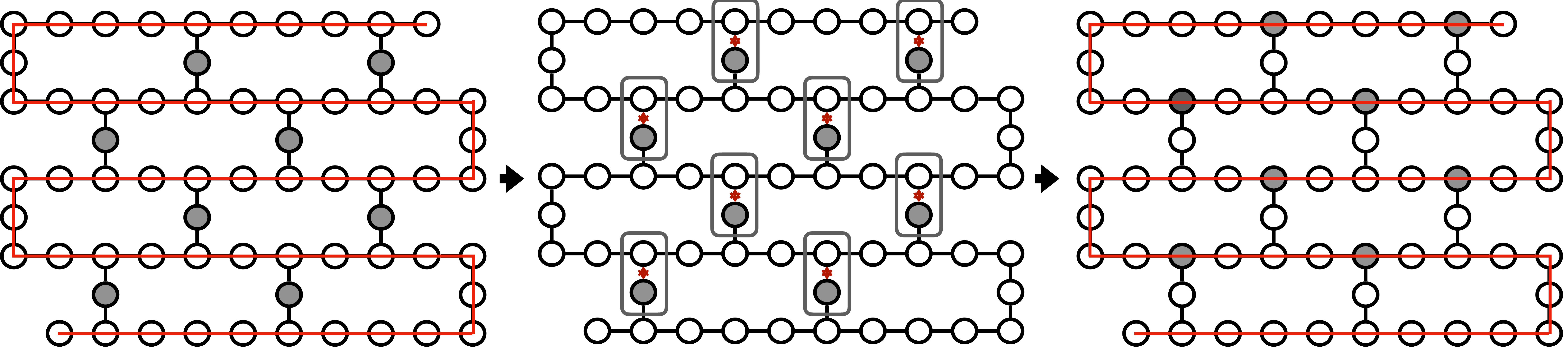}
    \caption{Laaps adapted to handle IBM Brooklyn and Washington types of architecture}
    \label{fig:ibm_pattern}
\end{figure*}
\begin{figure}[htb]
    \centering
    \includegraphics[width=0.4
    \textwidth]{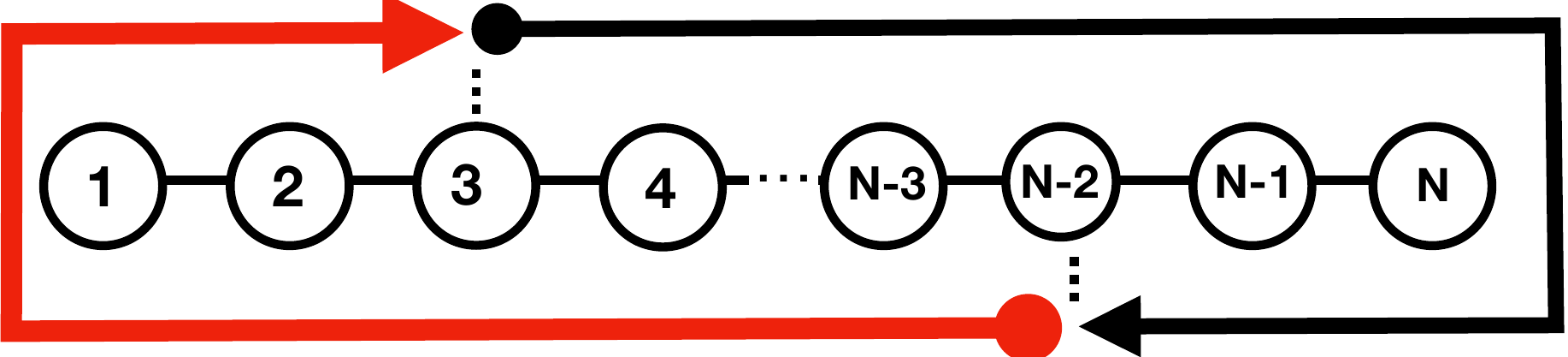}
    \caption{Two full iterations for the line architecture make each logical qubit visit every physical location.}
    \label{fig:ibm_2_iters}
\end{figure}

\vspace{-5pt}
\section{\textbf{Handling Real Problem Graphs}}
\label{sec:adapt}

Every problem graph of $n$ vertices is a subgraph of $n$-clique graph.
Since Section \ref{sec:basismethod} presents a method of executing all edges of a clique problem graph in linear time, any non-clique problem graph can also be completed in linear time.

However, in practice, the problem graph is usually not a complete graph.
Therefore, we further propose methods to adapt the Maaps algorithm to reduce the depth and gate count. We discuss two approaches: (1) Subgraph-isomorphism based, and (2) graph partition based. 

\subsection{The Subgraph-isomorphism Approach}

When we have a problem graph that is not a clique, we can try to avoid executing gates that do not appear in the problem graph, and then eliminate the cycles that do not include any active gate. We find the minimal cycle from which no more gate execution occurs. That will give us the actual number of cycles to execute a problem graph.

We find that mapping the nodes in the problem graph to the physical qubits in Maaps structure is important since different mapping will result in different depths.   We show an example in Fig.\ \ref{fig:initial_mapping_small}. 

\begin{figure*}[htb]
    \centering
    \includegraphics[width=0.8\textwidth]{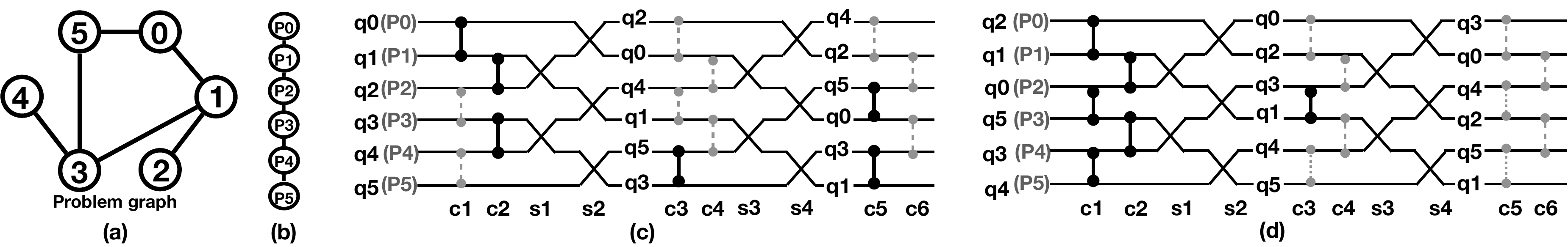}
    \caption{(a) Input problem graph. (b) A six-qubit linear physical coupling graph. (c) Following the line solution with one valid mapping, resulting in depth 8.  (d) Following the line solution with another valid mapping, resulting in depth 5.  The dashed-line gates in (c) and (d) are in original pattern but not in the input problem graph. The solid-line gates represents the gate in both.  }
    \label{fig:initial_mapping_small}
\end{figure*}

Finding the mapping from the problem graph to the Maaps    structure is important, yet challenging. The goal is to minimize the number of continuous cycles that cover active gates. We use a subgraph-isomorphism approach. 

Before we discuss this solution, we define a series of pattern graphs $pg\_k$,  $k = 1 ... n$ for Maaps   pattern. If we gradually build a graph based on which edges in the problem graph have been executed with respect to the computation cycle number, we will have a series of graphs. 



 Since our problem is to find the minimal number cycles, we can try to fit the problem graph into the each of the $pg\_k$ graphs, from k = 1 to $l$, where $l$ represents the first graph that can cover all edges in the problem graph. If the input problem graph is not a subgraph of $pg_k$, we advance to $pg_{k+1}$. This is guaranteed to terminate since any problem is a subgraph of the clique graph. At most we let $k$ be $n$ which is total number of vertices.

 Fitting a graph into another graph is a well studied problem, -- the subgraph-isomorphism problem. We use a recent subgraph-isomorphism algorithm  to implement it \cite{han+:icmd19}.
 
 


\begin{figure}[htb]
    \centering
    \includegraphics[width=0.5\textwidth]{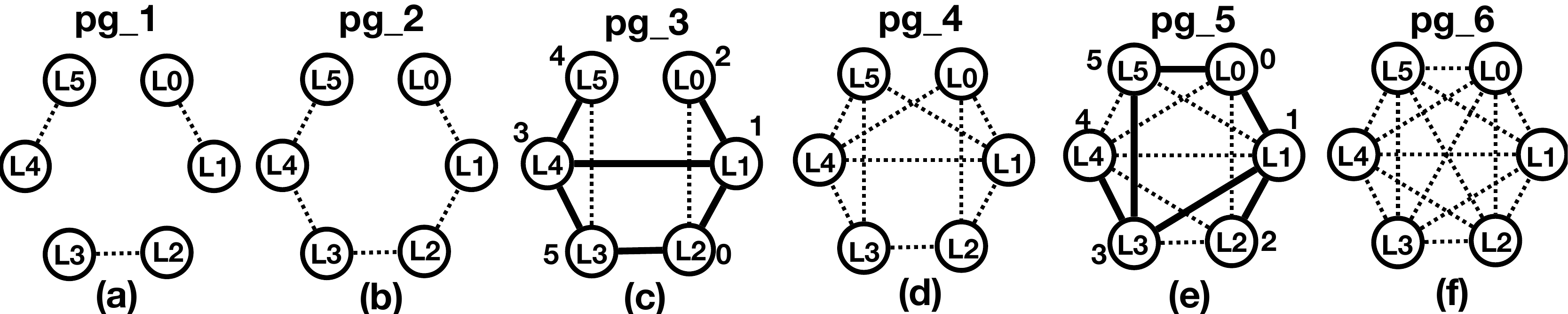}
    \caption{Pattern graphs for Laaps-6, corresponding to the execution sequence in Fig.\ \ref{fig:laaps_on_manila}. }
    \label{fig:pattern_graphs}
\end{figure}

\subsection{The graph partition approach}

However, the subgraph-isomorphism approach has large overhead. Our experiments show that it takes un-necessarily long time to find a mapping when the qubit number goes above 27 qubits. Moreover, it has to follow Laaps or Maaps exactly, which may not necessarily be the best case.  

When handling large problem graphs, we   partition a large problem graph into small but dense components and minimize the cut edges \cite{buluc+::jae16}\cite{karypis+:sc98}\cite{andreev+:tcs06}. 
The hardware coupling graph is also divided into multiple partitions such that each partition is mapped to one partition of the problem graph. 
 We define the degree $D_i$ of each partition $G_i$ as the number of cut-edges with one node in $G_i$.   We map the partition $G_i$ with highest $D_i$ to the center of hardware coupling graph. Then for each mapped graph partition $G_j$, we rank their un-mapped neighbours and pick the one with the highest degree to be mapped close to its mapped neighbour. We repeat this until all partitions are mapped. 

Within each hardware coupling partition, we use the Maaps  method. Different partitions run concurrently. After all partitions are handled separately, we handle the remaining cut edges. At this stage, the sub-graph with remaining edges will be sparse. 
We can divide them into multiple disjoint connected components and handle each one of them separately, again with Maaps    pattern. If it is very sparse, we can use a heuristic approach to move qubits that participate in two-qubit operators close to each other using the Floyd-Warshall algorithm, for one pair at one time.

{\paragraph{Performance guarantee:}} We acknowledge that introducing a heuristic method after graph partition might lose the performance guarantee. We tackle this problem for always keeping an upper-bound of the number of cycles during the compilation process. To start with, an initial mapping given by the graph partition method already implies how long it takes if we follow the exact Maaps    pattern. We save this bound. Then at each step, given an updated mapping, we  calculate how many more cycles based on following Maaps, and update the upper bound when necessary. In the end, if the compiled circuit has larger depth than the upper bound, we then revert to the point where this upper bound is achieved, and use it as the final compiled circuit. Hence it is still guaranteed to be better or the same as the linear depth we have defined in Section \ref{sec:basismethod}.

\section{Evaluation}
\label{sec:eval}

We evaluate our proposed methodology by comparing our results with state-of-the-art QAIM-IC \cite{alam+:micro20}, QAOA-OLSQ \cite{tan+:iccad20} and Paulihedral \cite{li+:asplos22}  with respect to depth, gate count, and fidelity. We performed experiments with three type of problem graphs: random graphs, regular graphs, and 2-local Hamitonian simulation graphs (for 1-D NN Ising model, 2-D NNN XY model, and 3-D NNN Heisenberg model). We vary vertex numbers  to up to 1024 and graph densities to up to 50\%. 


\subsection{{Experiment Setup}}

\noindent \textbf{Target Architectures} For large architecture, we use two popular architectures. One is the Google Sycamore which can be expanded to any number of qubits by adding diagonal connections. The original Sycamore structure is shown in Fig.\ \ref{fig:google_ibm}. The other is IBM Washington architecture in Fig.\ \ref{fig:google_ibm}, which comprises of connected 12-node rings. Every two adjacent rings share 3 qubits. It can be expanded to any number of 12-node rings. IBM Brooklyn and Washington (IBM's largest quantum machine) have such  structure. IBM Brooklyn has 8 complete rings while IBM Washington has 18 (or 16 due to temporarily down links )  complete rings.  

For small-scale experiments, we use synthetic 2D lattice architecture similar to that in \cite{alam+:micro20} and \cite{lao+:arxiv21twoqan}. We use four grid sizes: $3\times4$, $4\times4$, $4\times5$ and $5\times5$. \\


    
\noindent \textbf{Metrics} We use \textbf{depth} and \textbf{gate count} of the compiled circuit.  We also use the \textbf{estimated success probability} $(ESP)$ to evaluate the quality of the compiled circuits. $ESP$ is defined as the product of the success probability of individual gates. We also use the metric of \textbf{compilation time} to evaluate the efficiency of our compiler.\\

    

\noindent \textbf{Benchmarks}  We use the library Networkx in python \cite{networkx} to generate Erdös-Rényi random graphs and random $k$-regular graphs. For each type of graphs and each vertex number, we vary graph density to 0.3 and 0.5.


In addition, we also use the interaction graph for 2-local Hamiltonian, including 1-D NNN Ising model, 2-D NNN XY
model, and 3-D NNN Heisenberg model ('NNN' stands for the next nearest neighbouring) \cite{lao+:arxiv21twoqan}. \\

\noindent \textbf{{Baseline}} We compare our method with that by Alam \etal \cite{alam+:micro20}, which is named QAIM-IC, which is the fastest version among all proposed in their paper. QAOA-OLSQ \cite{tan+:iccad20}} which is an optimal scheduler for QAOA. We also compare our approach with  Paulihedral \cite{li+:asplos22}, reproduced by following the algorithm in the paper.


\begin{figure*}[ht]
     \centering
     \begin{subfigure}[b]{0.24\textwidth}
         \centering
         \includegraphics[width=\textwidth]{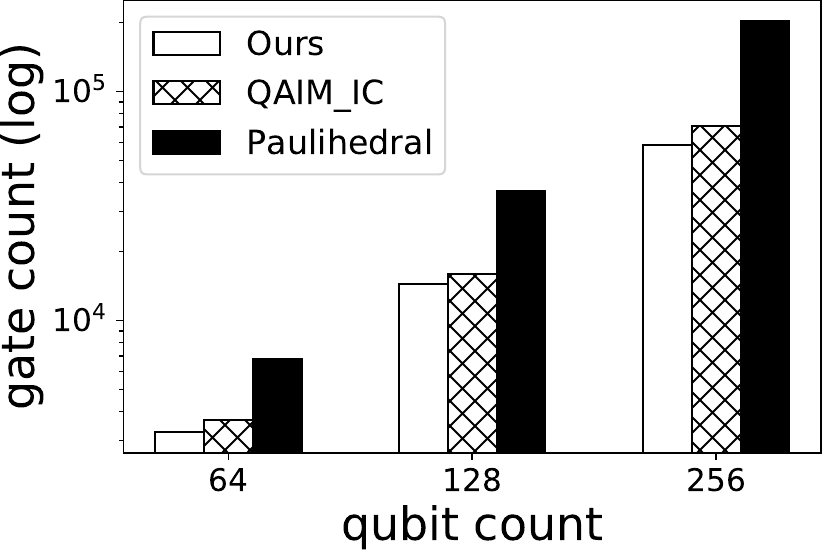}
         \caption{$Random\_5\_ibm$}
         \label{fig:GcRandom3ggl}
     \end{subfigure}
     \begin{subfigure}[b]{0.24\textwidth}
         \centering
         \includegraphics[width=\textwidth]{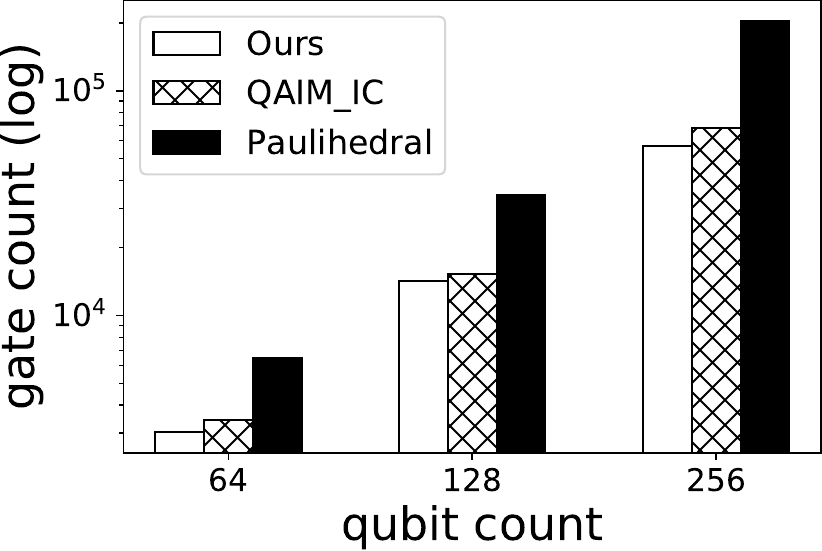}
         \caption{$Regular\_5\_ibm$}
         \label{fig:GcRandom5ggl}
     \end{subfigure}
     \begin{subfigure}[b]{0.24\textwidth}
         \centering
         \includegraphics[width=\textwidth]{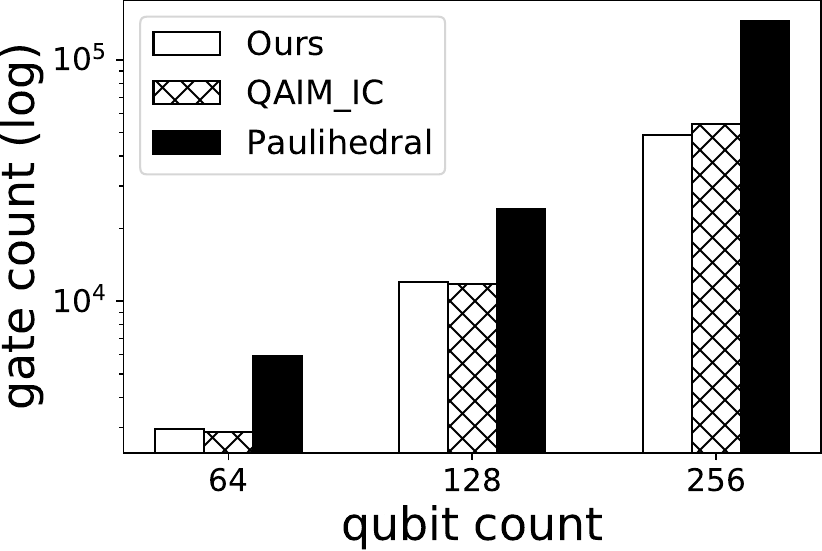}
         \caption{$Random\_5\_google$}
         \label{fig:GcRegular3ggl}
     \end{subfigure}
     \begin{subfigure}[b]{0.24\textwidth}
         \centering
         \includegraphics[width=\textwidth]{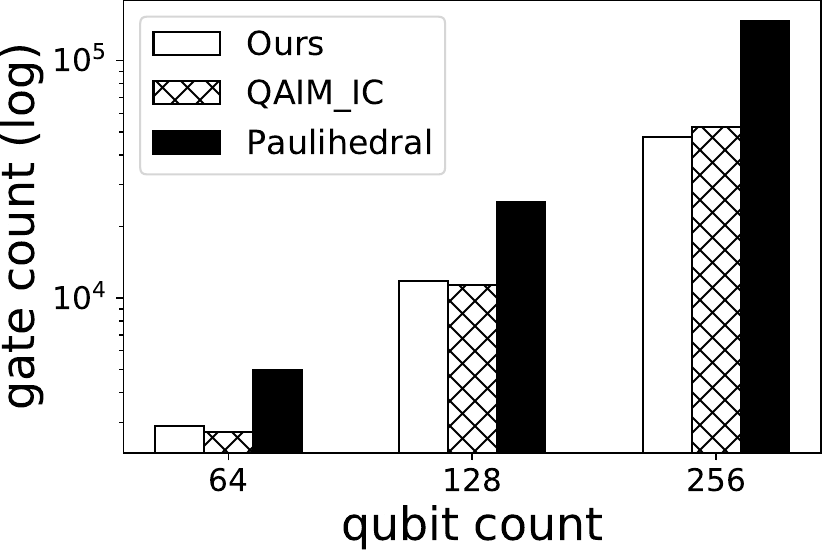}
         \caption{$Regular\_5\_google$}
         \label{fig:GcRegular5ggl}
     \end{subfigure}
        \caption{Gate count comparison}
        \label{fig:gc_figures}
\end{figure*}

\subsection{Impact on large-scale architectures}

 First, we show experiment results for the large scale problem graphs with different densities and qubit counts, on IBM heavy-hex and Google sycamore-like architectures.\\ 

\begin{figure*}[htb]
     \centering
     \begin{subfigure}[b]{0.24\textwidth}
         \centering
         \includegraphics[width=\textwidth]{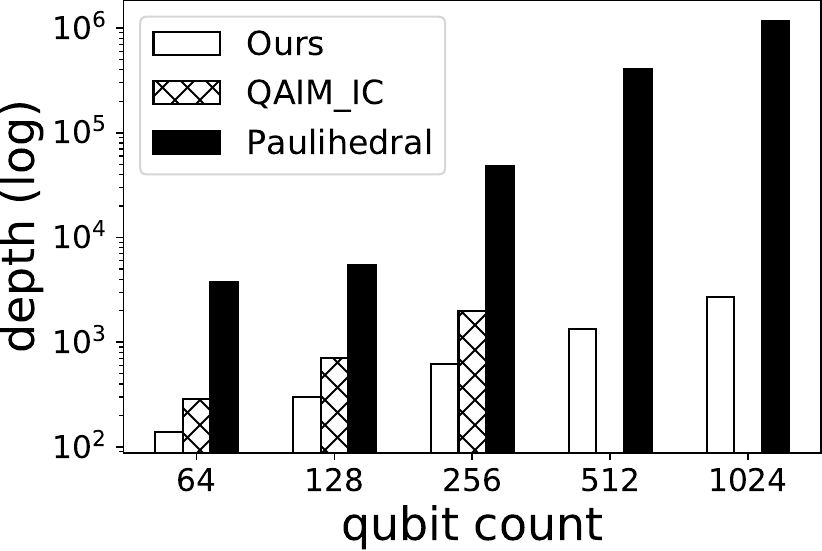}
         \caption{$Random\_3\_ibm$}
         \label{fig:Random3ibm}
     \end{subfigure}
     \begin{subfigure}[b]{0.24\textwidth}
         \centering
         \includegraphics[width=\textwidth]{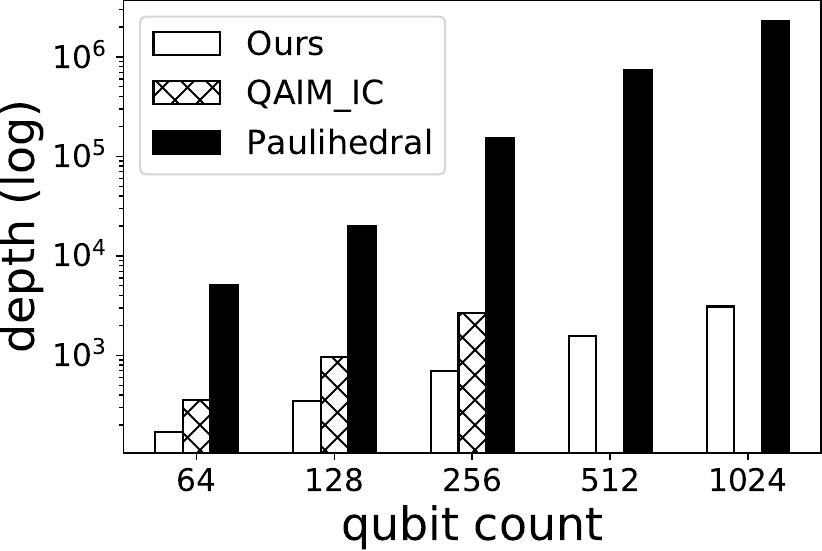}
         \caption{$Random\_5\_ibm$}
         \label{fig:Random5ibm}
     \end{subfigure}
     \begin{subfigure}[b]{0.24\textwidth}
         \centering
         \includegraphics[width=\textwidth]{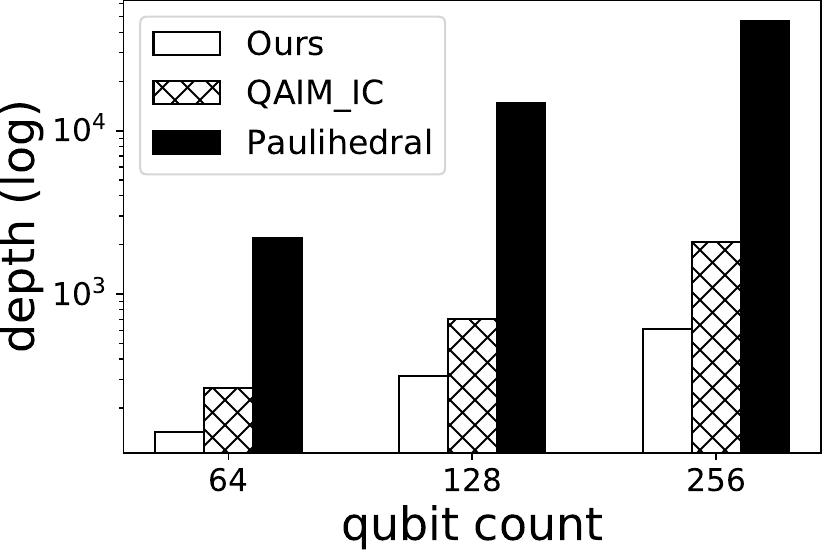}
         \caption{$Regular\_3\_ibm$}
         \label{fig:Regular3ibm}
     \end{subfigure}
     \begin{subfigure}[b]{0.24\textwidth}
         \centering
         \includegraphics[width=\textwidth]{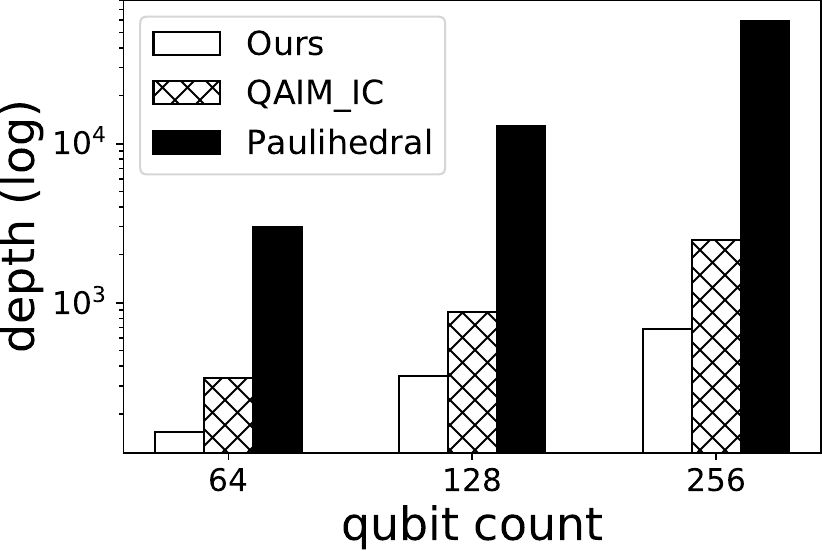}
         \caption{$Regular\_5\_ibm$}
         \label{fig:Regular5ibm}
     \end{subfigure}
        \caption{Depth comparison for IBM Architecture}
        \label{fig:ibm_bar_figures}
\end{figure*}

\begin{figure*}[htb]
     \centering
     \begin{subfigure}[b]{0.24\textwidth}
         \centering
         \includegraphics[width=\textwidth]{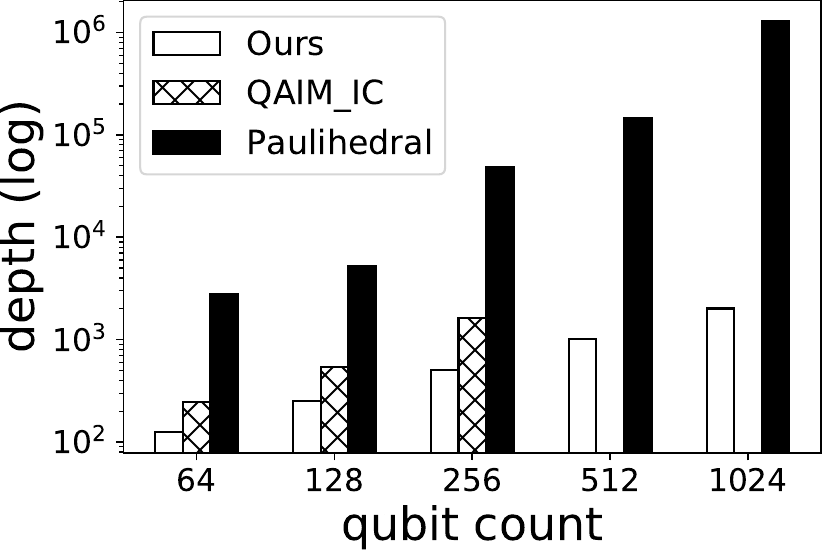}
         \caption{$Random\_3\_google$}
         \label{fig:Random3ggl}
     \end{subfigure}
     \begin{subfigure}[b]{0.24\textwidth}
         \centering
         \includegraphics[width=\textwidth]{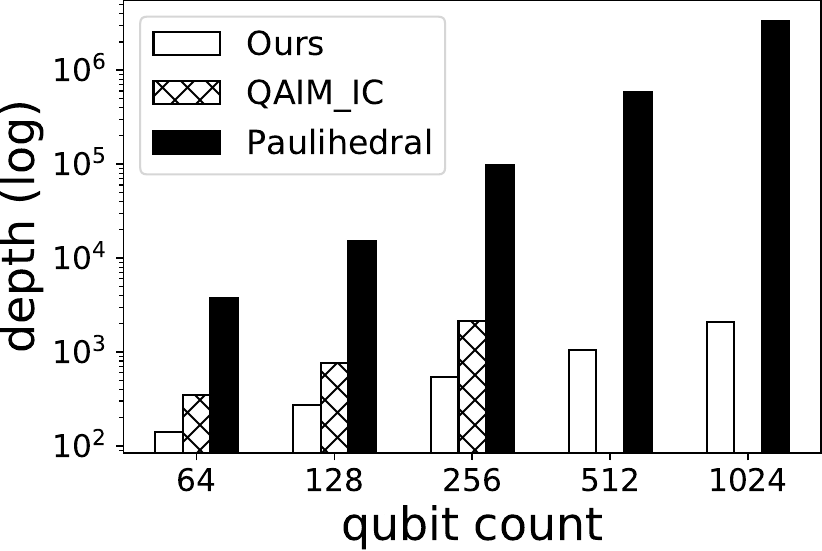}
         \caption{$Random\_5\_google$  }
         \label{fig:Random5ggl}
     \end{subfigure}
     \begin{subfigure}[b]{0.24\textwidth}
         \centering
         \includegraphics[width=\textwidth]{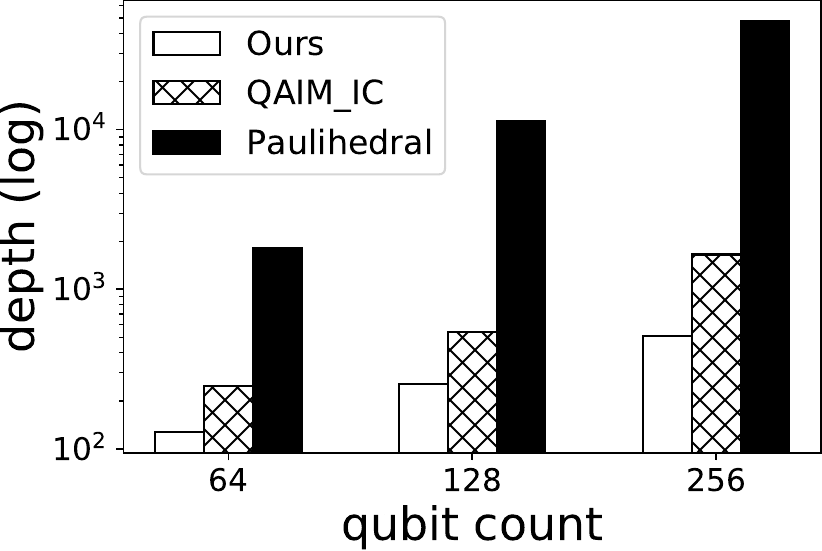}
         \caption{$Regular\_3\_google$}
         \label{fig:Regular3ggl}
     \end{subfigure}
     \begin{subfigure}[b]{0.24\textwidth}
         \centering
         \includegraphics[width=\textwidth]{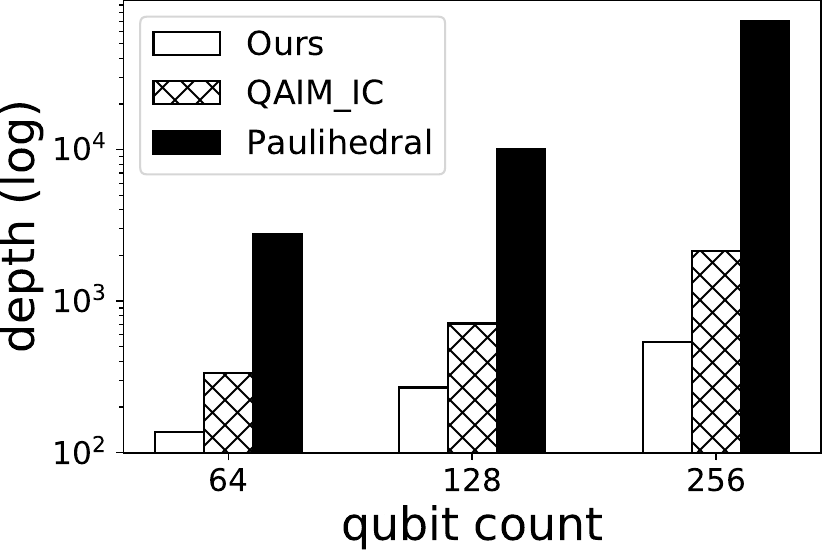}
         \caption{$Regular\_5\_google$}
         \label{fig:Regular5ggl}
     \end{subfigure}
        \caption{Depth comparison at Google}
        \label{fig:ggl_bar_figures}
\end{figure*}

\noindent \textbf{Depth}
The depths for the large scale experiments are shown at log-scale in Fig.\ \ref{fig:ibm_bar_figures} and \ref{fig:ggl_bar_figures}. In all experiments and on both architectures, our compiler significantly outperforms both baselines with respect to depth.

On the IBM heavy-hex architectures, our compiler achieves up to 3X improvement in depth over QAIM for random graphs with 0.3 density, and up to 4X improvement for random graphs with 0.5 density; the improvement over Paulihedral is even greater. Similar speedups are seen for the regular graphs.
In both cases, the improvement increases with the number of qubits, indicating that our compiler is able to scale better to larger problems compared to the baselines, thanks to the linear bound guaranteed by our approach.

On the Google sycamore-like architectures, 
we observe the same speedups over QAIM on both random and regular graphs.
Our compiler also achieves lower depths for each benchmark on the sycamore-like architecture, compared to on the heavy-hex architecture.
This is because for the IBM architectures, we ran the standard Laaps pattern on a line-embedding, whereas for the sycamore-like architecture we used the specialized Maaps pattern. The improvement demonstrates that our 2D Maaps pattern is able to take advantage of the increased connectivity of the non-linear sycamore-like architecture to improve upon the basic linear pattern.\\




\noindent \textbf{Gate Count}
The gate count results are shown at log-scale in Fig.\ \ref{fig:gc_figures}.
We only show results for experiments up to 256 qubits, as the QAIM compiler could not handle benchmark sizes larger than this.

Our compiler performs better than the Paulihedral baseline, and performs on par with the QAIM baseline. While our compiler achieves better gate counts on the IBM heavy-hex architecture, the QAIM compiler achieves better gate counts on the Google sycamore-like architectures the 64 and 128 qubit benchmarks.
This is because the Maaps pattern used for sycamore-like architectures uses many swaps to facilitate intra-row interactions, since there are no direct connections between qubits in a given row.
Despite this, our compiler still achieves gate counts very close to the QAIM compiler, and even achieves a better gate count on the sycamore-like architecture for 256 qubits. This suggests our compiler is able to scale to larger architectures better than QAIM with respect to gate count.

When calculating ESP on large benchmarks, the large number of gates quickly drives the product down to 0, so we do not directly show the numbers for large benchmarks. However, gate count provides a good indicator of ESP; a smaller gate count indicates higher ESP. Our gate counts are on par with or better than that of QAIM and Paulihedral. Decoherence error is related to depth; our compiled circuits have  significantly lower circuit depths, implying a much smaller decoherence error compared to QAIM.
Since fidelity combines these two factors of ESP and decoherence error, overall our compiler should achieve a better fidelity.\\


    
    
    
    

\noindent
\textbf{Compilation Time}
Compilation times for large benchmarks are shown in Table \ref{tab:comprisonTime}; the time is given in seconds. Our compiler runs much faster than the baseline. This is because, aside from the initial mapping of qubits, which can be done efficiently using existing graph partition methods, our compiler mainly follows a pre-defined pattern, which requires no intensive computation and only simple data structures. Thus the running time of our compiler depends on just the runtime of the graph partition as well as the length of the pattern, and our compiler is able to handle much larger benchmarks with ease, compared to the baseline.\\

\begin{table}[t]
\centering
\caption{Compilation time (s)}
\label{tab:comprisonTime}
\resizebox{0.45\textwidth}{!}{
\begin{tabular}{|c|c|cccc|}
\hline
  & Qubit count  & \multicolumn{1}{c|}{64}  & \multicolumn{1}{c|}{128} & \multicolumn{1}{c|}{512}  & \multicolumn{1}{c|}{1024}     
\\ 
\hline
\multicolumn{1}{|c|}{\textbf{Google}} & Ours     &  \multicolumn{1}{c|}{0.8}  & \multicolumn{1}{c|}{1.9} & \multicolumn{1}{c|}{6.6}  & \multicolumn{1}{c|}{22.3}      
\\ \cline{2-6}
\multicolumn{1}{|c|}{\textbf{Sycamore-like}} & QAIM     &  \multicolumn{1}{c|}{231}  & \multicolumn{1}{c|}{3479} & \multicolumn{1}{c|}{$>5hrs$}  & \multicolumn{1}{c|}{$>5hrs$}  
\\ \hline
\multicolumn{1}{|c|}{\textbf{IBM}} &Ours      & \multicolumn{1}{c|}{1.2} & \multicolumn{1}{c|}{6.4} & \multicolumn{1}{c|}{47.4} & \multicolumn{1}{c|}{336.6}      
\\  \cline{2-6}
\multicolumn{1}{|c|}{\textbf{heavy-hex}} &QAIM      & \multicolumn{1}{c|}{290} & \multicolumn{1}{c|}{4219} & \multicolumn{1}{c|}{$>5hrs$} & \multicolumn{1}{c|}{$>5hrs$}     
\\ \hline
\end{tabular}
}
\end{table}

\begin{figure}[h!]
 \centering
 \includegraphics[width=0.25\textwidth]{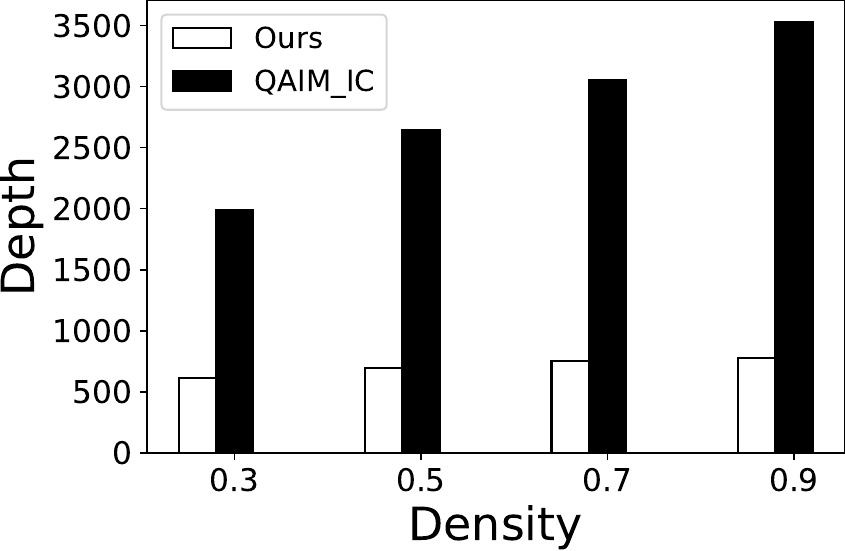}
 \caption{Depth of random graphs with 256 nodes and different densities for IBM architecture }
    
        \label{fig:random_x_ibm_depth}
\end{figure}

\noindent \textbf{Sensitivity to Density}
To explore the impact of density on performance, we fixed the number of qubits to 256, and ran experiments on the IBM heavy-hex architectures at various densities; the results are shown in Fig.\ \ref{fig:random_x_ibm_depth}.
While the depth achieved by the QAIM compiler increases from 2000 to 3500 as the density increases from 0.3 to 0.9, the depth achieved by our compiler only increases slightly from 600 to 800. We did not compare Paulihedral here as it is worse than QAIM\_IC.




\subsection{Impact on small-scale architectures}
We ran both random and regular graphs ranging from 10 to 25 qubits on small $N\times M$ 2D architectures. The results are shown in Table \ref{tab:SmallScaleComp}.

Our compilation achieves better depths than QAIM as well as better gate counts on all small benchmarks tested.
Since ESP is correlated to gate count, our compiler also achieves better ESP values for almost all the small benchmarks.
Unlike the larger benchmarks, where we used a heuristic based on graph partitioning to find an initial mapping, on smaller benchmarks we can afford to run a subgraph isomorphism algorithm to obtain the optimal mapping for following our Maaps pattern exactly.

{In addition, we also compare our approach with the  QAOA-OLSQ compiler \cite{tan+:iccad20}. The results are in Table.   \ref{tab:tanCompTan}. Due to the compilation overhead of QAOA-OLSQ, the comparison was only conducted for up to 15 qubits. 

 For the most of cases, our approach has better depth than QAOA-OLSQ, except one case 15-4, the 15-qubit density 40\% graph, where our depth is 11 and QAOA-OLSQ's depth is 9. However, in this case, OLSQ takes more than 2 days to run.  Our gate count is slightly worse than the QAOA-OLSQ method. But again, QAOA-OLSQ is a (near-)optimal solver and its compilation overhead is of magnitudes larger than ours. All these benchmarks are compiled within 0.3 seconds by our compiler. But it takes hours for QAOA-OLSQ when qubit number is beyond 15 and graph density is beyond 30\%.   
}


 

\begin{table}[t]
\centering
\caption{Small scale Comparison}
\label{tab:SmallScaleComp}
\resizebox{0.45\textwidth}{!}{
\begin{tabular}{|c|cc|cc|c|}
\hline
     & \multicolumn{2}{c|}{\textbf{Depth}} &\multicolumn{2}{c|}{\textbf{Gate Count}} & \multicolumn{1}{c|}{\textbf{ESP} }                       
\\ \hline
Graphs & \multicolumn{1}{c|}{Ours} & \multicolumn{1}{c|}{QAIM}& \multicolumn{1}{c|}{Ours} & \multicolumn{1}{c|}{QAIM} & \multicolumn{1}{c|}{Ours/QAIM}
\\ \hline
10-0.3  & \multicolumn{1}{c|}{6}  & \multicolumn{1}{c|}{11} & \multicolumn{1}{c|}{19}  & \multicolumn{1}{c|}{23}   & \multicolumn{1}{c|}{1.27X}
\\ \hline
15-0.3  & \multicolumn{1}{c|}{15}  & \multicolumn{1}{c|}{25} & \multicolumn{1}{c|}{64}  & \multicolumn{1}{c|}{65}   & \multicolumn{1}{c|}{1.06X}
\\ \hline
20-0.3  & \multicolumn{1}{c|}{22}  & \multicolumn{1}{c|}{33} & \multicolumn{1}{c|}{119}  & \multicolumn{1}{c|}{129}   & \multicolumn{1}{c|}{1.83X}
\\ \hline
25-0.3  & \multicolumn{1}{c|}{29}  & \multicolumn{1}{c|}{52} & \multicolumn{1}{c|}{194}  & \multicolumn{1}{c|}{215}   & \multicolumn{1}{c|}{3.57X}
\\ \hline
10-0.5  & \multicolumn{1}{c|}{15}  & \multicolumn{1}{c|}{22} & \multicolumn{1}{c|}{44}  & \multicolumn{1}{c|}{49}   & \multicolumn{1}{c|}{1.35X}
\\ \hline
15-0.5  & \multicolumn{1}{c|}{17}  & \multicolumn{1}{c|}{32} & \multicolumn{1}{c|}{87}  & \multicolumn{1}{c|}{95}   & \multicolumn{1}{c|}{1.62X}
\\ \hline
20-0.5  & \multicolumn{1}{c|}{24}  & \multicolumn{1}{c|}{53} & \multicolumn{1}{c|}{159}  & \multicolumn{1}{c|}{190}   & \multicolumn{1}{c|}{6.54X}
\\ \hline
25-0.5  & \multicolumn{1}{c|}{34}  & \multicolumn{1}{c|}{66} & \multicolumn{1}{c|}{259}  & \multicolumn{1}{c|}{307}   & \multicolumn{1}{c|}{18.34X}
\\ \hline

10-3  & \multicolumn{1}{c|}{6}  & \multicolumn{1}{c|}{14} & \multicolumn{1}{c|}{24}  & \multicolumn{1}{c|}{28}   & \multicolumn{1}{c|}{1.27X}
\\ \hline
10-4  & \multicolumn{1}{c|}{10}  & \multicolumn{1}{c|}{13} & \multicolumn{1}{c|}{33}  & \multicolumn{1}{c|}{34}   & \multicolumn{1}{c|}{1.06X}
\\ \hline
15-4  & \multicolumn{1}{c|}{11}  & \multicolumn{1}{c|}{19} & \multicolumn{1}{c|}{57}  & \multicolumn{1}{c|}{60}   & \multicolumn{1}{c|}{1.20X}
\\ \hline
15-6  & \multicolumn{1}{c|}{16}  & \multicolumn{1}{c|}{34} & \multicolumn{1}{c|}{79}  & \multicolumn{1}{c|}{83}   & \multicolumn{1}{c|}{1.27X}
\\ \hline
20-6  & \multicolumn{1}{c|}{17}  & \multicolumn{1}{c|}{28} & \multicolumn{1}{c|}{107}  & \multicolumn{1}{c|}{113}   & \multicolumn{1}{c|}{1.44X}
\\ \hline
20-8  & \multicolumn{1}{c|}{22}  & \multicolumn{1}{c|}{36} & \multicolumn{1}{c|}{142}  & \multicolumn{1}{c|}{155}   & \multicolumn{1}{c|}{2.20X}
\\ \hline
25-8  & \multicolumn{1}{c|}{28}  & \multicolumn{1}{c|}{47} & \multicolumn{1}{c|}{209}  & \multicolumn{1}{c|}{219}   & \multicolumn{1}{c|}{1.83X}
\\ \hline
25-10  & \multicolumn{1}{c|}{30}  & \multicolumn{1}{c|}{71} & \multicolumn{1}{c|}{235}  & \multicolumn{1}{c|}{279}   & \multicolumn{1}{c|}{14.39X}
\\ \hline
\end{tabular}
}
\end{table}

\begin{table}[t]
{
\centering
\caption{{Comparison with QAOA-OLSQ}}
\label{tab:tanCompTan}
\resizebox{0.45\textwidth}{!}{
\begin{tabular}{|c|cc|cc|cc|}
\hline
     & \multicolumn{2}{c|}{\textbf{Depth}} &\multicolumn{2}{c|}{\textbf{Gate Count}} & \multicolumn{2}{c|}{\textbf{Compilation time(s)} }                       
\\ \hline
Graphs & \multicolumn{1}{c|}{Ours} & \multicolumn{1}{c|}{qaoa-olsq}& \multicolumn{1}{c|}{Ours} & \multicolumn{1}{c|}{qaoa-olsq} & \multicolumn{1}{c|}{Ours} & \multicolumn{1}{c|}{qaoa-olsq} 
\\ \hline 
10-2  & \multicolumn{1}{c|}{4}  & \multicolumn{1}{c|}{5} & \multicolumn{1}{c|}{12}  & \multicolumn{1}{c|}{10}   & \multicolumn{1}{c|}{0.001}&
\multicolumn{1}{c|}{0.24}
\\ \hline
10-3  & \multicolumn{1}{c|}{6}  & \multicolumn{1}{c|}{7} & \multicolumn{1}{c|}{22}  & \multicolumn{1}{c|}{19}   & \multicolumn{1}{c|}{0.003}&
\multicolumn{1}{c|}{33.2}
\\ \hline
10-4  & \multicolumn{1}{c|}{11}  & \multicolumn{1}{c|}{12} & \multicolumn{1}{c|}{32}  & \multicolumn{1}{c|}{25}   & \multicolumn{1}{c|}{0.04}&
\multicolumn{1}{c|}{283.4}
\\ \hline
12-2  & \multicolumn{1}{c|}{3}  & \multicolumn{1}{c|}{4} & \multicolumn{1}{c|}{16}  & \multicolumn{1}{c|}{12}   & \multicolumn{1}{c|}{0.006}&
\multicolumn{1}{c|}{0.32}
\\ \hline
12-3  & \multicolumn{1}{c|}{8}  & \multicolumn{1}{c|}{7} & \multicolumn{1}{c|}{26}  & \multicolumn{1}{c|}{22}   & \multicolumn{1}{c|}{0.09}&
\multicolumn{1}{c|}{4100.16}
\\ \hline
12-4  & \multicolumn{1}{c|}{11}  & \multicolumn{1}{c|}{12} & \multicolumn{1}{c|}{41}  & \multicolumn{1}{c|}{30}   & \multicolumn{1}{c|}{0.07}&
\multicolumn{1}{c|}{17141.2}
\\ \hline
15-2  & \multicolumn{1}{c|}{4}  & \multicolumn{1}{c|}{6} & \multicolumn{1}{c|}{19}  & \multicolumn{1}{c|}{16}   & \multicolumn{1}{c|}{0.26}&
\multicolumn{1}{c|}{640.5}
\\ \hline
15-4  & \multicolumn{1}{c|}{11}  & \multicolumn{1}{c|}{9} & \multicolumn{1}{c|}{50}  & \multicolumn{1}{c|}{40}   & \multicolumn{1}{c|}{0.04}&
\multicolumn{1}{c|}{>2 days}
\\ \hline
\end{tabular}
}}

\end{table}									

\subsection{2-local Hamiltonian problem graphs}
The results for 2-local Hamiltonian experiments are shown in Table \ref{tab:2local}.
These were run on medium scale 64-qubit Google sycamore-like and IBM ring-life architectures.
Our compiler beats the QAIM baseline for all 2-local benchmarks.

\begin{table}[t]
\centering
\caption{Depth Comparison for 2-Local Hamtiltonian}
\label{tab:2local}
\resizebox{0.45\textwidth}{!}{
\begin{tabular}{|c|c|c|c|c|}
\hline
 &   & \multicolumn{1}{c|}{1D Ising}& \multicolumn{1}{c|}{2D XY}& \multicolumn{1}{c|}{3D Heisenberg}
 \\ \hline
\multicolumn{1}{|c|}{\textbf{Google}} & \multicolumn{1}{c|}{Ours}   & \multicolumn{1}{c|}{40}  & \multicolumn{1}{c|}{91}  & \multicolumn{1}{c|}{103 }                    
\\ \cline{2-5}
\multicolumn{1}{|c|}{\textbf{Sycamore-64}} & \multicolumn{1}{c|}{QAIM} & \multicolumn{1}{c|}{96} & \multicolumn{1}{c|}{384} & \multicolumn{1}{c|}{510} 
 \\ \hline
 \multicolumn{1}{|c|}{\textbf{IBM}} & \multicolumn{1}{c|}{Ours}   & \multicolumn{1}{c|}{86}  & \multicolumn{1}{c|}{123}  & \multicolumn{1}{c|}{104 }                                     
\\ \cline{2-5}
 \multicolumn{1}{|c|}{\textbf{heavy-hex-64}}& \multicolumn{1}{c|}{QAIM} & \multicolumn{1}{c|}{87} & \multicolumn{1}{c|}{423} & \multicolumn{1}{c|}{483} 
 \\ \hline

\end{tabular}
}
\end{table}

\section{Related Work}
\label{sec:rel}
Existing compilation approaches include generic compilers that assume the gates in a circuit have a fixed set of partial order. There has been extensive study in generic compilers \cite{zhang+:asplos21, Li+:ASPLOS19,murali+:asplos19,siraichi+:oopsla19,Siraichi+:CGO18,Zulehner+:DATE18,Zulehner+:ICRC17, tan+:iccad20, tannu+:asplos19, Wille+:DAC19, Shafaei+:ASPDAC14}.

QAOA \cite{QAOA:farhi2014quantum, QAOA:farhi2017quantum, QAOA:farhi2021quantum} however is a special type of application that allows commutativity among two-qubit operators. It exposes new optimization opportunity for generating better depth and fidelity circuits. Hence there has been a handful of studies for compiling and optimizing QAOA \cite{lao+:arxiv20, alam+:micro20, tan+:iccad20, li+:asplos22, lao+:arxiv21twoqan}. Tan \etal \cite{tan+:iccad20} provide a constrained SAT solver to achieve optimality. Alam \etal \cite{alam+:micro20} use the heuristic of connectivity strength. Paulihedral \cite{li+:asplos22} handles generic Pauli-strings in Hamiltionian simulation. 2QAN \cite{lao+:arxiv21twoqan} proposes an algorithm that has quadratic-time complexity and uses unitary unifying to enhance the results. In particular, 2QAN not only handles QAOA circuit but also 2-local Hamiltonian simulation circuits.

But none of these studies provide performance guarantee or rigorous theoretical analysis as to how far the obtained circuit is from the optimal for multi-dimensional architectures. The consequence is that it may generate circuit with un-necessarily large depth and hence sabotage the depth advantage of QAOA or Hamiltonian simulation. Our study is the first that provides a structured method with performance guarantee for multi-dimensional architecture together with a concurrent work by Weidenfeller \etal \cite{weidenfeller+:arxiv22}. It has low compilation overhead and hence scales to large architectures.

\section{Conclusion}
\label{sec:ccl}
We present a novel pattern-based compiler framework  for QAOA and 2-local Hamiltonian simulation circuit with linear bound. The pattern applies to multi-dimensional architecture. We demonstrate how to adapt it to Google Sycamore and IBM architectures. Our methodology excels in depth and gate count for both large and small scale benchmarks.


\clearpage
\bibliographystyle{IEEEtranS}
\bibliography{refs}


\end{document}